\documentclass[12pt]{article}
\usepackage{epsfig}

\font\blackboard=msbm10 at 12pt
\font\blackboards=msbm7
\font\blackboardss=msbm5
\newfam\black
\textfont\black=\blackboard
\scriptfont\black=\blackboards
\scriptscriptfont\black=\blackboardss
\def\bb#1{{\fam\black\relax#1}}

\newcommand{\junk}[1]{}

\newcommand{\ba}{\begin{array}}
\newcommand{\ea}{\end{array}}
\newcommand{\be}{\begin{equation}}
\newcommand{\ee}{\end{equation}}
\newcommand{\bea}{\begin{eqnarray}}
\newcommand{\eea}{\end{eqnarray}}
\newcommand{\beas}{\begin{eqnarray}}
\newcommand{\eeas}{\end{eqnarray}}

\def\identity{{\rlap{1} \hskip 1.6pt \hbox{1}}}

\def\laplace{{\kern1pt\vbox{\hrule height 1.2pt\hbox{\vrule width
1.2pt\hskip
  3pt\vbox{\vskip 6pt}\hskip 3pt\vrule width 0.6pt}\hrule height
  0.6pt}
  \kern1pt}}
\def\scriptlap{{\kern1pt\vbox{\hrule height 0.8pt\hbox{\vrule width
  0.8pt
  \hskip2pt\vbox{\vskip 4pt}\hskip 2pt\vrule width 0.4pt}\hrule height
  0.4pt}
  \kern1pt}}

\def\roughly#1{\raise.3ex\hbox{$#1$\kern-
.75em\lower1ex\hbox{$\sim$}}}

\newcommand{\NP}{{\em Nucl.\ Phys.\ }}

\newcommand{\PL}{{\em Phys.\ Lett.\ }}
\newcommand{\PR}{{\em Phys.\ Rev.\ }}
\newcommand{\PRP}{{\em Phys.\ Rep.\ }}

\newcommand{\MPL}{{\em Mod.\ Phys.\ Lett.\ }}
\newcommand{\PRL}{{\em Phys.\ Rev.\ Lett.\ }}

\newcommand{\gone}[1]{}
\begin{document}
\pagestyle{plain}
\setcounter{page}{1}

\baselineskip16pt

\begin{titlepage}

\begin{flushright}
MIT-CTP-3341\\
hep-th/0301094
\end{flushright}
\vspace{13 mm}

\begin{center}

{\Large \bf Lectures on D-branes, tachyon condensation, and string
  field theory
\\}
\vspace{3mm}

\end{center}
\vspace{0.2in}

\begin{center}
Notes on lectures presented at the CECS School on Quantum Gravity,\\
Valdivia, Chile, January 2002

\end{center}

\vspace{7 mm}

\begin{center}

Washington Taylor

\vspace{3mm}
{\small \sl Center for Theoretical Physics} \\
{\small \sl MIT, Bldg.  6-308} \\
{\small \sl Cambridge, MA 02139, U.S.A.} \\
{\small \tt wati@mit.edu}\\
\end{center}

\vspace{8 mm}

\begin{abstract}
These lectures provide an introduction to the subject of tachyon
condensation in the open bosonic string.  The problem of tachyon
condensation is first described in the context of the low-energy
Yang-Mills description of a system of multiple D-branes, and then
using the language of string field theory.  An introduction is given
to Witten's cubic open bosonic string field theory.  The Sen
conjectures on tachyon condensation in open bosonic string field
theory are introduced, and evidence confirming these conjectures is
reviewed.
\end{abstract}

\vspace{1cm}
\begin{flushleft}
November 2002
\end{flushleft}
\end{titlepage}
\newpage


\section{Introduction}
\label{sec:introduction}

The last seven years have been a very exciting time for string theory.
A new understanding of nonperturbative features of string theory, such
as D-branes, has led to exciting new developments relating string
theory to physically interesting systems such as black holes and
supersymmetric gauge field theories, as well as to a new understanding
of the relationship between Yang-Mills theories and quantum theories
of gravity.

Despite remarkable progress in these directions, however, a consistent
nonperturbative background-independent formulation of string theory is
still lacking.  This situation makes it impossible at this point, even
in principle, to directly address cosmological questions using string
theory.  String field theory is a nonperturbative approach to string
theory which holds some promise towards providing
a background-independent definition
of the theory.  These lecture notes give an introduction to string
field theory and review some recent work which incorporates D-branes
into the framework of string field theory.  This
work shows that string field theory is a sufficiently robust framework
that distinct string backgrounds can arise as disconnected solutions
of the theory, at least for open strings.  It remains to be seen
whether this success can be replicated in the closed string sector.

In this section we review briefly the situation in string theory as a
whole, and summarize the goals of this set of lectures.  In Section 2
we review some basic aspects of D-branes.  In Section 3, we describe a
particular D-brane configuration which exhibits a tachyonic
instability.  This tachyon can be seen in the low-energy super
Yang-Mills description of the D-brane geometry.  
This field theory tachyon provides a simple model which embodies much
of the physics of the more complicated string field theory tachyon
discussed in the later lectures.
In Section 4 we
give an introduction to Witten's cubic
bosonic open string field theory and summarize the
conjectures made by Sen in 1999, which suggested
that the tachyonic instability of the open bosonic string can be
interpreted in terms of an unstable space-filling D-brane, and that
this system can be analytically described through open string field
theory.  Section 5 gives a more
detailed analytic description of Witten's cubic string field theory.
In Section  6 we
summarize evidence from string field theory for Sen's conjectures.  
Section 7 contains a brief review of some more
recent developments.  Section 8
contains concluding remarks and lists some open problems.

Much new work has been done in this area since these lectures were
presented at Valdivia in January 2002.  Except for a few references to
more recent developments in footnotes and in the last two sections,
these lecture notes primarily cover work done before January 2002.
Previous articles reviewing related work include those of Ohmori
 \cite{Ohmori}, de Smet \cite{deSmet}, and Aref'eva {\it et al.} \cite{abgkm}.
An expanded set of lecture notes, based on lectures given by the
 author and Barton Zwiebach at TASI '01, will appear in
 \cite{Taylor-Zwiebach}; the larger set of notes will include further
 details on a number of topics.

\subsection{The status of string theory: a brief review}
\label{sec:status}

To understand the significance of developments over the last seven
years, it is useful to recall the situation of string theory as it was
in early 1995.  At that time it was clearly understood that there were
5 distinct ways in which a supersymmetric closed string could be
quantized to give a microscopic definition  of a theory of quantum
gravity in ten dimensions.  Each of these approaches to quantizing the
string gives a set of rules for calculating scattering amplitudes
between on-shell string states which describe gravitational quanta as
well as an infinite family of massive particles in a ten-dimensional
spacetime.  These five string theories are known as the type IIA, IIB,
I, heterotic $SO(32)$, and heterotic $E_8 \times E_8$ superstring
theories.  While these string theories give perturbative descriptions
of quantum gravity, in 1995 there was little understanding of
nonperturbative aspects of these theories.

In the years between 1995 and 2000, several new ideas dramatically
transformed our understanding of string theory.  We now
briefly summarize these ideas and mention some aspects of these
developments relevant to the main topic of these lectures.
\vspace*{0.07in}

\noindent 
{\bf Dualities:} The five different perturbative formulations of
superstring theory are all related to one another through duality
symmetries \cite{Hull-Townsend,Witten-various}, whereby  the
degrees of freedom in one theory can be described through a duality
transformation in terms of the degrees of freedom of another theory.
Some of these duality symmetries are nonperturbative, in the sense
that the string coupling $g$ in one theory is related to the inverse
string coupling $1/g$ in a dual theory.  The web of dualities relating
the different theories gives a picture in which, rather than
describing five distinct possibilities for a fundamental theory, each
of the perturbative superstring theories appears to be a particular
perturbative limit of some other, as yet unknown, underlying
theoretical structure.
\vspace*{0.07in}

\noindent 
{\bf M-theory:} In addition to the five perturbative string theories,
the web of dualities also seems to include a limit which describes a
quantum theory of gravity in eleven dimensions.  This new theory has
been dubbed ``M-theory''.  Although no covariant definition for
M-theory has been given, this theory can be related to type IIA and
heterotic $E_8 \times E_8$ string theories through compactification on
a circle $S^1$ and the space $S^1/Z_2$
respectively \cite{Townsend-11,Witten-various,Horava-Witten}.  For example, in
relating to the type IIA theory, the compactification radius $R_{11}$
of M-theory becomes the product $ gl_s$ of the string coupling and
string length in the 10D IIA theory.  Thus, M-theory in flat space,
which arises in the limit $R_{11} \rightarrow \infty$, can be thought
of as the strong coupling limit of type IIA string theory.  It is also
suspected that M-theory may be describable as a quantum theory of
membranes in 11 dimensions \cite{Townsend-11}, although a covariant
formulation of such a theory is still lacking.
\vspace*{0.07in}

\noindent 
{\bf Branes:} In addition to strings, all five superstring
theories, as well as M-theory, contain extended objects of higher
dimensionality known as ``branes''.  M-theory has M2-branes and
M5-branes, which have two and five dimensions of spatial extent
(whereas a string has one).  The different superstring theories each
have different complements of D-branes as well as the fundamental
string and Neveu-Schwarz 5-brane; in particular, the IIA/IIB
superstring theories contain D-branes of all even/odd dimensions.  The
branes of one theory can be related to the branes of another through
the duality transformations mentioned above.  Through an appropriate
sequence of dualities, any brane can be mapped to  any other
brane, including the string itself.  This suggests that none of these
objects are really any more fundamental than any others; this idea is
known as ``brane democracy''.
\vspace*{0.07in}

\noindent 
{\bf M(atrix) theory and AdS/CFT:} One of the most remarkable results
of the developments just mentioned is the realization that in certain
space-time backgrounds, M-theory and string theory can be completely
described through simple supersymmetric quantum mechanics and field
theory models related to the low-energy description of systems of
branes.  The M(atrix) model of M-theory is a simple supersymmetric
matrix quantum mechanics which is believed to capture all of the
physics of M-theory in asymptotically flat spacetime (in light-cone
coordinates).  A closely related set of higher-dimensional
supersymmetric Yang-Mills theories are related to string theory in
backgrounds described by the product of anti-de Sitter space and a
sphere through the AdS/CFT correspondence.  It is believed that these
models of M-theory and string theory give true nonperturbative
descriptions of quantum gravity in space-time backgrounds which have
the asymptotic geometry relevant to each model.  For reviews of
M(atrix) theory and AdS/CFT, see \cite{WT-RMP,agmoo}.
\vspace*{0.1in}

The set of ideas just summarized have greatly increased our
understanding of nonperturbative aspects of string theory.  In
particular, through M(atrix) theory and the AdS/CFT correspondences we
now have nonperturbative definitions of M-theory and string theory in
certain asymptotic space-time backgrounds which could, in principle,
be used to calculate any local result in quantum gravity.  While these
new insights are very powerful, however, we are still lacking a truly
background-independent formulation of string theory.

\subsection{The goal of these lectures}

The goal of these lectures is to describe progress towards a
nonperturbative background-independent formulation of string theory.
Such a formulation is needed to address fundamental questions such as:
What is string theory/M-theory?  How is the vacuum of string theory
selected?  ({\it i.e.}, Why can the observable low-energy universe be
accurately described by the standard model of particle physics in four
space-time dimensions with an apparently small but nonzero positive
cosmological constant?), and other questions of a cosmological nature.
Obviously, aspiring to address these questions is an ambitious
undertaking, but we believe that attaining a better understanding of
string field theory is a useful step in this direction.

More concretely, in these lectures we will describe recent progress on
open string field theory. It may be useful here to recall some basic
aspects of open and closed strings and the relationship between them.

Closed strings, which are topologically equivalent to a circle $S^1$,
give rise upon quantization to a massless set of spacetime fields associated
with the graviton $g_{\mu \nu}$, the dilaton $\varphi$, and the
antisymmetric two-form $B_{\mu \nu}$, as well as an infinite family of
massive fields.  For the supersymmetric closed string, further
massless fields associated with the graviton supermultiplet
appear---these are the Ramond-Ramond $p$-form fields $A^{(p)}_{\mu_1
\cdots \mu_p}$ and the gravitini $\psi_{\mu \alpha}$.  Thus, the
quantum theory of closed strings is naturally associated with a theory
of gravity in space-time.  On the other hand, open strings, which are
topologically equivalent to an interval $[0, \pi]$, give rise under
quantization to a massless gauge field $A_\mu$ in space-time.  The
supersymmetric open string also has a massless gaugino field
$\psi_\alpha$.  It is now understood that generally open strings
should be thought of as ending on a Dirichlet $p$-brane (D$p$-brane),
and that the massless open string fields describe the fluctuations of
the D-brane and the gauge field living on the world-volume of the D-brane.

It may seem, therefore, that open and closed strings are quite
distinct, and describe disjoint aspects of the physics in a fixed
background space-time containing some family of D-branes.  At tree
level, the closed strings indeed describe gravitational physics in the
bulk space-time, while the open strings describe the D-brane dynamics.
At the quantum level, however, the physics of open and closed strings
are deeply connected.  Indeed, historically open strings were
discovered first through the form of their scattering amplitudes
 \cite{Veneziano}.  Looking at one-loop processes for open strings led
to the first discovery of closed strings, which appeared as {\em
poles} in nonplanar one-loop open string diagrams \cite{NGS,Lovelace}.  The
fact that open string diagrams naturally contain closed string
intermediate states indicates that in some sense all closed string
interactions are implicitly defined through the complete set of open
string diagrams.  This connection underlies many of the important
recent developments in string theory.  In particular, the M(atrix)
theory and AdS/CFT correspondences between gauge theories and quantum
gravity are essentially limits in which closed string physics in a
fixed space-time background is captured by a simple limiting
Yang-Mills description of an open string theory on a family of branes
(D0-branes for M(atrix) theory, D3-branes for the CFT describing
AdS${}_5\times S^5$, etc.)

The fact that, in certain fixed space-time backgrounds, quantum
gravity theories can be encoded in terms of open string degrees of
freedom through the M(atrix) and AdS/CFT correspondences leads to the
question of how a global change of the space-time background would
appear in the quantum field theory describing the appropriate limit of
the open string model in question.  If such a change of background
could be described in the context of M(atrix) theory or AdS/CFT, it
would indicate that these models could be generalized to a
background-independent framework.  Unfortunately, however, such a
change in the background involves adding nonrenormalizable
interactions to the field theories in question.  At this point in time
we do not have the technology to understand generically how a sensible
quantum field theory can be described when an infinite number of
nonrenormalizable interaction terms are added to the Lagrangian.  One
example of a special case where this can be done is the addition of a
constant background $B$ field in space-time.  In the associated
Yang-Mills theory, such as that on a system of $N$ D3-branes in the
case of the simplest AdS/CFT correspondence, this change in the
background field corresponds to replacing products of open string
fields with a noncommutative star-product.  The resulting theory is a
noncommutative Yang-Mills theory.  Such noncommutative theories are
the only well-understood example of a situation where adding an
infinite number of apparently nonrenormalizable terms to a field
theory action leads to a sensible modification of quantum field
theory (for a review of noncommutative field theory and its connection
to string theory, see \cite{Douglas-Nekrasov}).

String field theory is a nonperturbative formulation in target space
of an interacting string theory, in which the infinite family of
fields associated with string excitations are described by a
space-time field theory action. For open strings, this field theory is
a natural extension of the low-energy Yang-Mills action describing a
system of D-branes, where the entire hierarchy of massive string
fields is included in addition to the massless gauge field on the
D-brane.  Integrating out all the massive fields from the string field
theory action gives rise to a nonabelian Born-Infeld action for the
D-branes, including an infinite set of higher-order terms arising from
string theory corrections to the simple Yang-Mills action.  Like the
case of noncommutative field theory discussed above, the new terms
appearing in this action are apparently nonrenormalizable, but the
combination of terms must work together to form a sensible theory.

In the 1980's, a great deal of work was done on formulating string
field theory for open and closed, bosonic and supersymmetric string
theories.  Most of these string field theories are quite complicated.
For the open bosonic string, however, Witten \cite{Witten-SFT}
constructed an extremely elegant string field theory based on the
Chern-Simons action.  This cubic bosonic open string field theory
(OSFT) is the primary focus of the work described in these lectures.
Although Witten's OSFT can be described in a simple abstract language,
practical computations with this theory rapidly become extremely
complicated.  Despite a substantial amount of work on this theory,
little insight was gained in the 1980's regarding how this theory
could be used to go beyond standard perturbative string methods.  Work
on this subject stalled out in the late 80's, and little further
attention was paid to OSFT until several years ago.

One simple feature of the 26-dimensional bosonic string has been
problematic since the early days of string theory: both the open and
closed bosonic strings have tachyons in their spectra, indicating that
the usual perturbative vacua used for these
theories are unstable.  In 1999, Ashoke Sen had a remarkable insight
into the nature of the open bosonic string
tachyon \cite{Sen-universality}.  He observed that the open bosonic
string should be thought of as ending on a space-filling D25-brane.
He pointed out that this D-brane is unstable in the bosonic theory, as
it does not carry any conserved charge,
and he suggested that the open bosonic string tachyon should be
interpreted as the instability mode of the D25-brane.  This led him to
conjecture that Witten's open string field theory could be used to
precisely determine a new vacuum for the open string, namely one in
which the D25-brane is annihilated through condensation of the
tachyonic unstable mode.  Sen made several precise conjectures
regarding the details of the string field theory description of this
new open string vacuum.  As we describe in these lectures, there is
now overwhelming evidence that Sen's picture is correct, demonstrating
that string field theory accurately describes the nonperturbative
physics of D-branes.  This new nonperturbative application of string
field theory has sparked a new wave of work on Witten's cubic open
string field theory, revealing many remarkable new structures.  
In particular, string field theory now provides a concrete framework
in which disconnected string backgrounds can emerge from the equations
of motion of a single underlying theory.  Although so far this can
only be shown explicitly in the open string context, this work paves
the way for a deeper understanding of background-independence in
quantum theories of gravity.


\section{D-branes}
\label{sec:D-branes}

In this section we briefly review some basic features of D-branes.
The concepts developed here will be useful in describing tachyonic
D-brane configurations in the following section.  For more detailed
reviews of D-branes, see \cite{Polchinski-TASI,WT-Trieste}.

\subsection{D-branes and Ramond-Ramond charges}

D-branes can be understood in two ways:  {\it a}) as extended extremal
black brane solutions of supergravity carrying conserved charges, and
{\it b}) as hypersurfaces on which strings have Dirichlet boundary
conditions.
\vspace*{0.03in}

\noindent
{\it a}) The ten-dimensional type IIA and IIB supergravity theories
each have a set of $(p + 1)$-form fields $A^{(p + 1)}_{\mu_1 \cdots
\mu_{(p+ 1)}}$ in the supergraviton multiplet, with $p$ even/odd for
type IIA/IIB supergravity.  These are the Ramond-Ramond fields in the
massless superstring spectrum.  For each of these $(p + 1)$-form
fields, there is a solution of the supergravity field equations which
has $(p + 1)$-dimensional Lorentz invariance, and which has the form
of an extremal black hole solution in the orthogonal $9-p$ space
directions plus time (for a review see \cite{dkl}).  These ``black
$p$-brane'' solutions carry charge under the R-R fields $A^{(p + 1)}$,
and are BPS states in the supergravity theory, preserving half the
supersymmetries of the theory.
\vspace*{0.03in}

\noindent
{\it b})
In type IIA and IIB string theory, it is possible to consider open
strings with Dirichlet boundary conditions on some number $9-p$ of
the spatial coordinates $x^\mu (\sigma)$.  The locus of points defined
by such Dirichlet boundary conditions defines a $(p + 1)$-dimensional
hypersurface $\Sigma_{p + 1}$ in the ten-dimensional spacetime.
When $p$ is even/odd in
type IIA/IIB string theory, the spectrum of the resulting quantum open
string theory contains a massless set of fields $A_\alpha, \alpha = 0,
1, \ldots, p$ and $X^{a}, a = p + 1, \ldots, 9$.
These fields can be associated with a gauge field living on the
hypersurface $\Sigma_{p + 1}$, and a set of  degrees of
freedom describing the transverse fluctuations of this hypersurface in
spacetime.  Thus, the quantum fluctuations of the open string describe
a fluctuating $(p + 1)$-dimensional hypersurface in spacetime --- a
Dirichlet-brane, or ``D-brane''.

The remarkable insight of Polchinski in 1995 \cite{Polchinski}
was the observation that Dirichlet-branes carry Ramond-Ramond
charges, and therefore should be described in the low-energy
supergravity limit of string theory by precisely the black $p$-branes
discussed in {\it a}).  This connection between the string and
supergravity descriptions of these nonperturbative objects paved the
way to a dramatic series of new developments in string theory,
including connections between string theory and supersymmetric gauge
theories, string constructions of black holes, and new approaches to
string phenomenology.

\subsection{Born-Infeld and super Yang-Mills D-brane actions}

In this subsection we briefly review the low-energy super Yang-Mills
description of the dynamics of one or more D-branes.  As discussed in
the previous subsection, the massless open string modes on a
D$p$-brane in type IIA or IIB superstring theory describe a $(p +
1)$-component gauge field $A_\alpha$, $9-p$ transverse scalar fields
$X^a$, and a set of massless fermionic gaugino fields.  The scalar
fields $X^a$ describe small fluctuations of the D-brane around a flat
hypersurface.  If the D-brane geometry is sufficiently far from flat,
it is useful to describe the D-brane configuration by a general
embedding $X^\mu (\xi)$, where $\xi^\alpha$ are $p + 1$ coordinates on
the D$p$-brane world-volume $\Sigma_{(p + 1)}$, and $X^\mu$ are ten
functions giving a map from $\Sigma_{(p + 1)}$ into the space-time
manifold ${\bb R}^{ 9, 1}$.  Just as the Einstein equations governing
the geometry of spacetime arise from the condition that the one-loop
contribution to the closed string beta function vanish, a set of
equations of motion for a general D$p$-brane geometry and associated
world-volume gauge field can be derived from a calculation of the
one-loop open string beta function \cite{Leigh}.  These equations of
motion arise from the classical Born-Infeld action
\begin{equation}
S = - T_p  \int d^{p + 1} \xi
\;e^{-\varphi} \;\sqrt{-\det (G_{\alpha \beta} + B_{\alpha \beta} + 2 \pi \alpha'
F_{\alpha \beta})  }   + S_{{\rm CS}}+{\rm fermions}
\label{eq:DBI}
\end{equation}
where $G$, $B$  and $\varphi$ are the pullbacks of the 10D metric,
antisymmetric tensor and dilaton to the D-brane world-volume, while $F$ is the
field strength of the world-volume $U(1)$ gauge field $A_{\alpha}$.
$S_{\rm CS}$ represents a set of Chern-Simons terms which will be
discussed in the following subsection.
This action can be verified by a perturbative string
calculation \cite{Polchinski-TASI}, which also gives a precise
expression for the brane tension
\begin{equation}
\tau_p =\frac{T_p}{g} =
 \frac{1}{g\sqrt{\alpha'}}  \frac{1}{ (2 \pi \sqrt{\alpha'})^{p}} 
\end{equation}
where $g = e^{\langle \varphi \rangle}$ is the string coupling, equal to
the exponential 
of the dilaton expectation value.

A particular limit of the Born-Infeld action
(\ref{eq:DBI}) is useful for describing many low-energy aspects of
D-brane dynamics.  Take the background space-time $G_{\mu \nu}=
\eta_{\mu \nu}$ to be flat, and all other supergravity fields ($B_{\mu
\nu}, A^{(p + 1)}_{\mu_1 \cdots \mu_{p + 1}}$) to vanish.  We then
assume that the D-brane is approximately flat, and is close to the
hypersurface $X^a = 0, a > p$, so that we may make the static gauge
choice $X^\alpha = \xi^\alpha$.  We furthermore assume that
$\partial_\alpha X^a$ and $2 \pi \alpha' F_{\alpha \beta}$ are small
and of the same order.  In this limit, the action (\ref{eq:DBI})
can be expanded as
\begin{equation}
S =-\tau_pV_p  -\frac{1}{4g_{{\rm YM}}^2}
 \int d^{p + 1} \xi
\left(F_{\alpha \beta} F^{\alpha \beta} +\frac{2}{(2 \pi \alpha')^2} 
\partial_\alpha X^a \partial^\alpha X^a\right) +\cdots
\label{eq:action-expansion}
\end{equation}
where $V_p$ is the $p$-brane world-volume and the coupling $g_{{\rm
YM}}$ is given by
\begin{equation}
g_{{\rm YM}}^2 = \frac{1}{4 \pi^2 \alpha'^2 \tau_p} 
= \frac{g}{\sqrt{\alpha'}}  (2 \pi \sqrt{\alpha'})^{p-2} \,.
\label{eq:ym-coupling}
\end{equation}
Including fermionic terms, the second term in
(\ref{eq:action-expansion}) is simply the
dimensional reduction to $(p + 1)$ dimensions of the 10D 
${\cal N} = 1 $ super Yang-Mills action
\begin{equation}
S = \frac{1}{g_{{\rm YM}}^2}  \int d^{10}\xi \; \left(
 -\frac{1}{4} F_{\mu \nu}F^{\mu \nu}
+ \frac{i}{2}  \bar{\psi} \Gamma^\mu \partial_{\mu} \psi \right)
\label{eq:SYM1}
\end{equation}
where for $\alpha, \beta \leq p$, $F_{\alpha \beta}$ is the
world-volume U(1) field strength, and for $a > p, \alpha \leq p$, $
F_{\alpha a} \rightarrow \partial_\alpha X^a$
(setting $2 \pi \alpha' = 1$).

When multiple D$p$-branes are present, the D-brane action is modified
in a fairly simple fashion \cite{Witten-multiple}.  Consider a system
of $N$ coincident D-branes.  For every pair of branes $\{i, j\}$ there
is a set of massless fields
\begin{equation}
(A_\alpha)_i^{\; j},  \; \; \;(X^a)_i^{\; j}
\label{eq:nonabelian-fields}
\end{equation}  
associated with
strings stretching from the $i$th brane to the $j$th brane; the
indices $i, j$ are known as Chan-Paton indices.  Treating the fields
(\ref{eq:nonabelian-fields}) as matrices, the analogue for multiple
branes of the Born-Infeld action (\ref{eq:DBI}) takes the form
\begin{equation}
S \sim \int {\rm Tr}\; \sqrt{-\det \left( G + B + F \right)}\,.
\label{eq:NBI}
\end{equation}
This action is known as the nonabelian Born-Infeld action (NBI).  In
order to give a rigorous definition to the nonabelian Born-Infeld
action, it is necessary to resolve ordering ambiguities in the
expression (\ref{eq:NBI}).  Since the spacetime coordinates $X^a$
associated with the D-brane positions in space-time become themselves
matrix-valued, even evaluating the pullbacks $G_{\alpha \beta},
B_{\alpha \beta}$ involves resolving ordering issues.  Much work has
been done recently to resolve these ordering ambiguities (see
 \cite{ordering-NBI} for some recent papers in this direction which
contain further references to the literature), but there is still no
consistent definition of the nonabelian Born-Infeld theory
(\ref{eq:NBI}) which is valid to all orders.

The nonabelian Born-Infeld action (\ref{eq:NBI}) becomes much simpler
in the low-energy limit when the background space-time is flat.  In
the same limit discussed above for the single D-brane, where we find a
low-energy limit giving the U(1) super Yang-Mills theory in $p + 1$
dimensions, the inclusion of multiple D-branes simply leads in the
low-energy limit to the nonabelian U(N) super Yang-Mills action in $p
+ 1$ dimensions.  This action is the dimensional reduction of the 10D
U(N) super Yang-Mills action (analogous to (\ref{eq:SYM1}), but with
an overall trace) to $p + 1$ dimensions.
In this reduction, as before,
 for $\alpha, \beta \leq p$, $F_{\alpha \beta}$ is the
world-volume U(1) field strength, and for $a > p, \alpha \leq p$, $
F_{\alpha a} \rightarrow \partial_\alpha X^a$, where now $ A_\alpha,
X^a,$ and $F_{\alpha \beta}$ are $N \times N$ matrices.  We
furthermore have, for $a, b > p$, $F_{ab} \rightarrow -i[X^a, X^b]$ in
the dimensional reduction.  

The low-energy description of a system of $N$ coincident flat D-branes
is thus given by $U(N)$ super Yang-Mills theory in the appropriate
dimension.  This connection between D-brane actions in string theory
and super Yang-Mills theory has led to many new developments,
including new insights into supersymmetric field theories, the
M(atrix) theory and AdS/CFT correspondences, and brane world
scenarios.

\subsection{Branes from branes}

In this subsection we describe a remarkable feature of D-brane
systems, namely a mechanism by which one or more D-branes of a fixed
dimension can be used to construct additional D-branes of higher or
lower dimension.

In our discussion of the D-brane action (\ref{eq:DBI}) above, we
mentioned a group of terms $S_{\rm CS}$ which we did not describe
explicitly.  For a single D$p$-brane, these Chern-Simons terms can be
combined into a single expression of the form
\begin{equation}
S_{\rm CS}\sim \int_{\Sigma_{p + 1}}\;{\cal A} \;e^{F + B}
\label{eq:Chern-Simons}
\end{equation}
where ${\cal A} = \sum_{k}A^{(k)} $ represents a formal sum over all
the Ramond-Ramond fields $A^{(k)}$ of various dimensions.  In this
integral, for each term $A^{(k)}$, the nonvanishing contribution to
(\ref{eq:Chern-Simons}) is given by expanding the exponential of $F +
B$ to order $(p + 1-k)/2$, where the dimension of the resulting form
saturates the dimension of the brane.  For example, on  a D$p$-brane,
there is a coupling of the form
\begin{equation}
\int_{\Sigma_{(p + 1)}} A^{(p-1)} \wedge F\,.
\end{equation}
This coupling implies that the U(1) field strength on the D$p$-brane
couples to the R-R field associated with $(p-2)$-branes.  Thus, we can
associate magnetic fields on a D$p$-brane with dissolved
$(p-2)$-branes living on the D$p$-brane.  This result generalizes to a
system of multiple D$p$-branes by simply performing a trace on the RHS
of (\ref{eq:Chern-Simons})
For example, on $N$ compact
D$p$-branes, the charge 
\begin{equation}
 \frac{1}{2 \pi}  \int {\rm Tr}\; F_{\alpha \beta},
\label{eq:p-2-charge}
\end{equation}
which is the first Chern class of the U(N) bundle 
described by the gauge field on the $N$ branes,
is quantized and measures the number of units of D$(p-2)$-brane charge
living on the D$p$-branes, which are encoded in the field strength
$F_{\alpha \beta}$.  Similarly,
\begin{equation}
 \frac{1}{8 \pi^2}  \int {\rm Tr}\; F\wedge F
\end{equation}
encodes D$(p -4)$-brane charge on the D$p$-branes.

Just as lower-dimensional branes can be described in terms of the
degrees of freedom associated with a system of $N$ D$p$-branes through
the field strength $F_{\alpha \beta}$, higher-dimensional branes can
be described by a system of $N$ D$p$-branes in terms of the
commutators of the matrix-valued scalar fields $X^a$.  Just as
$\frac{1}{2 \pi} F$ measures $(p-2)$-brane charge, the matrix 
\begin{equation}
2 \pi i
[X^a, X^b]
\label{eq:up-charge}
\end{equation} measures $ (p + 2)$-brane charge
 \cite{WT-Trieste,Mark-Wati-4,Myers}.  The charge (\ref{eq:up-charge})
should be interpreted as a form of local charge density.
The fact that the
trace of (\ref{eq:up-charge}) vanishes for finite sized matrices
corresponds to the fact that the net D$p$-brane charge of a
finite-size brane configuration in flat spacetime vanishes.  

A simple example of the mechanism by which a system of multiple
D$p$-branes form a higher-dimensional brane is given by the matrix
sphere.  If we take a system of D0-branes with scalar matrices $X^a$
given by
\begin{equation}
X^a = \frac{2r}{ N}  J^a, \;\;\;\;\; a = 1, 2, 3
\label{eq:matrix-sphere}
\end{equation}
where $J^a$ are the generators of SU(2) in the $N$-dimensional
representation, then we have a configuration corresponding to the
``matrix sphere''.  This is a D2-brane of spherical geometry living on
the locus of points satisfying $x^2 + y^2 + z^2 = r^2$.  The ``local''
D2-brane charge of this brane is given by (\ref{eq:up-charge}).
The D2-brane configuration given by (\ref{eq:matrix-sphere}) is
rotationally invariant (up to a gauge transformation).  The
restriction of the brane to the desired locus of
points  can be seen from the relation $(X^1)^2 + (X^2)^2 + (X^3)^2 =
r^2\identity+{\cal O} (N^{-2})$.

\subsection{T-duality}

We conclude our discussion of D-branes with a brief description of
T-duality.  T-duality is a perturbative symmetry which relates the
type IIA and type IIB string theories.  This duality symmetry was in
fact crucial in the original discovery of D-branes \cite{Polchinski}.
A more detailed discussion of T-duality can be found in the textbook
by Polchinski \cite{Polchinski-string}.
Using T-duality, we construct an explicit example of a brane within a
brane encoded in super Yang-Mills theory, illustrating the ideas of
the previous subsection.  This example will be used in the following
section to construct an analogous configuration with a tachyon.

Consider type IIA string theory on a spacetime of the form $M^9 \times
S^1$ where $M^9$ is a generic 9-manifold of Lorentz signature, and
$S^1$ is a circle of radius $R$.  T-duality is the statement that this
theory is precisely equivalent, at the perturbative level, to type IIB
string theory on the spacetime $M^9 \times (S^1)'$, where
$(S^1)'$ is a circle of radius $R' = \alpha'/R$.

T-duality is most easily understood in terms of closed strings, where
it amounts to an exchange of winding and momentum modes of the string.
The string winding modes on $S^1$ have energy $m = R w/\alpha'$, where
$w$ is the winding number.  the T-dual momentum modes on $(S^1)'$ have
$m = n/R'$; it is straightforward to check that the spectrum of closed
string states is unchanged under T-duality.  T-duality can also be
understood in terms of open strings.  Under T-duality, an open string
with Neumann boundary conditions on $S^1$ is mapped to an open string
with Dirichlet boundary conditions on $(S^1)'$, and vice versa.  Thus,
a Dirichlet $p$-brane which is wrapped around the circle $S^1$ is
mapped under T-duality to a Dirichlet $(p-1)$-brane of one lower
dimension which is localized to a point on the circle $(S^1)'$.  At
the level of the low-energy theory on the D-brane, the $(p +
1)$-dimensional Yang-Mills theory on the $p$-brane is replaced under
T-duality with the $p$-dimensional Yang-Mills theory on the dual
$(p-1)$-brane.  Mathematically, the covariant derivative operator in
the direction $S^1$ is replaced under T-duality with an adjoint scalar
field $X^a$.  Formally, this adjoint scalar field is an infinite size
matrix, containing information about the open strings wrapped an
arbitrary number of times around the compact direction $(S^1)'$.

We can summarize the relevant mappings under T-duality in the
following table

\begin{center}
\begin{tabular}{rcr}
IIA/$S^1$ & $\leftrightarrow $ & IIB/$(S^1)'$\\
$R$ & $\leftrightarrow $ & $R' =\alpha'/R $\\
Neumann/Dirichlet b.c.'s & $\leftrightarrow $ & Dirichlet/Neumann b.c.'s\\
$p$-brane & $\leftrightarrow $ & $(p\pm 1)$-brane\\
$2 \pi \alpha' (i \partial_a + A_a)$ & $\leftrightarrow $ & $X^a $
\end{tabular}
\end{center}
\vspace*{0.2in}

The phenomena by which field strengths in one brane describe lower- or
higher-dimensional branes can be easily understood using T-duality.
The following simple example may help to clarify this connection.
(For a more detailed discussion from this point of view see
 \cite{WT-Trieste}.) 
\begin{figure}
\epsfig{file=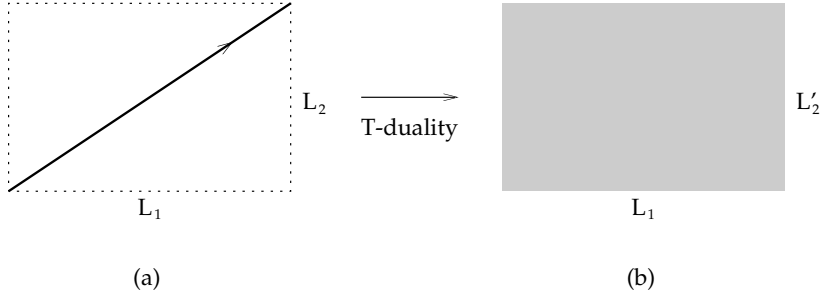,width=15cm}
\caption[x]{\footnotesize T-duality takes a diagonal D1-brane  on a
  two-torus (a) to a D2-brane on the dual torus with constant magnetic
  flux encoding an embedded D0-brane (b).}
\label{f:T-duality}
\end{figure}

Consider a D1-brane wrapped diagonally on a two-torus $T^2$ with sides
of length $L_1 = L$ and $L_2 = 2 \pi R$.
(Figure~\ref{f:T-duality}(a)).   This configuration is described in
terms of the world-volume Yang-Mills theory on a D1-brane stretched in
the $L_1$ direction through a transverse scalar field
\begin{equation}
X^2 = 2 \pi R \xi_1/L\,.
\end{equation}
To be technically precise, this scalar field should be treated as an
$\infty \times \infty$ matrix \cite{WT-T-duality} whose $(n, m)$ entry
is associated with strings connecting the $n$th and $m$th images of the
D1-brane on the covering space of $S^1$.  The diagonal elements
$X^2_{n,n}$ of this infinite matrix are given by $2 \pi R (\xi_1 +
n L)/L$, while all off-diagonal elements vanish.  While the resulting
matrix-valued function of $\xi_1$ is not periodic, it is periodic up
to a gauge transformation
\begin{equation}
X^2 (L) = V X^2 (0) V^{-1}
\label{eq:boundary-1}
\end{equation}
where $V$ is the shift matrix with nonzero elements $V_{n, n + 1} = 1$.

Under T-duality
in the $x^2$ direction
the infinite matrix $X^2_{nm}$ becomes the Fourier mode representation of a
gauge field on a dual D2-brane
\begin{equation}
A_2 = \frac{1}{ R' L}  \xi_1\,.
\end{equation}
The magnetic flux associated with this gauge field is
\begin{equation}
F_{12} = \frac{1}{ R' L} 
\end{equation}
so that
\begin{equation}
\frac{1}{2 \pi} \int F_{12} \; d \xi^1 \, d \xi^2 = 1\,.
\label{eq:0-charge}
\end{equation}
Note that the boundary condition (\ref{eq:boundary-1}) on the
infinite matrix $X^2$ transforms under T-duality to the boundary
condition on the gauge field
\begin{eqnarray}
A_2 (L, x_2) & = &  
e^{2 \pi i \xi_2/L_2'}
\left(A_2 (0, x_2)  + i \partial_2 \right)
e^{-2 \pi i \xi_2/L_2'}\\
& = &e^{2 \pi i \xi_2/L_2'}
A_2 (0, x_2) 
e^{-2 \pi i \xi_2/L_2'} + \frac{2 \pi}{ L_2'}, \nonumber
\end{eqnarray}
where the off-diagonal elements of the shift matrix $V$ in
(\ref{eq:boundary-1}) describe winding modes which correspond after
T-duality to the first Fourier mode $e^{2 \pi i \xi_2/L_2'}$.  The
boundary condition on the gauge fields in the $\xi_2$ direction is
trivial, which simplifies the T-duality map; a similar construction
can be done with a nontrivial boundary condition in both directions,
although the configuration looks more complicated in the D1-brane
picture.

This construction gives a simple Yang-Mills description of the mapping
of D-brane charges under T-duality: the initial configuration
described above has charges associated with a single D1-brane wrapped
around each of the directions of the 2-torus: D$1_1 +$ D$1_2$.  Under
T-duality, these D1-branes are mapped to a D2-brane and a D0-brane
respectively: D$2_{12} +$ D$0$.  The flux integral (\ref{eq:0-charge})
is the representation in the D2-brane world-volume Yang-Mills theory
of the charge associated with a D0-brane which has been uniformly
distributed over the surface of the D2-brane, just as in
(\ref{eq:p-2-charge}).


\section{Tachyons and D-branes}
\label{sec:tachyon-D-branes}

We now turn to the subject of tachyons.  Certain D-brane
configurations are unstable, both in supersymmetric and
nonsupersymmetric string theories.  This instability is manifested as
a tachyon with $M^2 < 0$ in the spectrum of open strings ending on the
D-brane.  We will explicitly describe the tachyonic mode in the case
of the open bosonic string in Section \ref{sec:bosonic-string}; this
open bosonic string tachyon will be the focal point of most of the
developments described in these notes.  In this section we list some
elementary D-brane configurations where tachyons arise, and we
describe a particular situation in which the tachyon can be seen in
the low-energy Yang-Mills description of the D-branes.  This
Yang-Mills background with a tachyon provides a simple field-theory
model of a system analogous to the more complicated string field
theory tachyon we describe in the later part of these notes.  This
simpler model may be useful to keep in mind in the later analysis.

\subsection{D-brane configurations with tachyonic instabilities}
\label{sec:D-brane-tachyons}

Some simple examples of unstable D-brane configurations where the open
string contains a tachyon include the following:
\vspace*{0.1in}

{\bf Brane-antibrane:} A pair of parallel D$p$-branes with opposite
orientation in type IIA or IIB string theory which are separated by a
distance $d < l_s$ give rise to a tachyon in the spectrum of open
strings stretched between the branes \cite{Banks-Susskind}.  The
difference in orientation of the branes means that the two branes are
really a brane and antibrane, carrying equal but opposite R-R charges.
Since the net R-R charge is 0, the brane and antibrane can annihilate,
leaving an uncharged vacuum configuration.
\vspace*{0.05in}

{\bf Wrong-dimension branes:} In type IIA/IIB string theory, a
D$p$-brane of even/odd spatial dimension $p$ is a stable BPS state
carrying a nonzero R-R charge.  On the other hand, a D$p$-brane of the
{\it wrong} dimension ({\it i.e.,} odd/even for IIA/IIB) carries no
charges under the classical IIA/IIB supergravity fields, and has a
tachyon in the open string spectrum.  Such a brane can annihilate to
the vacuum without violating charge conservation.
\vspace*{0.05in}

{\bf Bosonic D-branes:} Like the wrong-dimension branes of IIA/IIB
string theory, a D$p$-brane of any dimension in the bosonic string
theory carries no conserved charge and has a tachyon in the open
string spectrum.  Again, such a brane can annihilate to the vacuum
without violating charge conservation.

\subsection{Example: tachyon in low-energy field theory of two D-branes}
\label{sec:example-SYM}

As an example of how tachyonic configurations behave physically, we
consider in this subsection a simple example where a brane-antibrane
tachyon can be seen in the context of the low-energy Yang-Mills
theory.  This system was originally considered in
 \cite{Hashimoto-Taylor,gns}.  

\begin{figure}
\epsfig{file=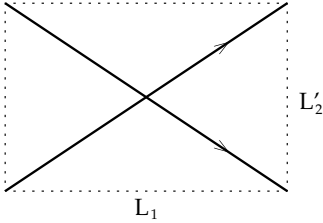,width=15cm}
\caption[x]{\footnotesize A pair of crossed D1-branes, T-dual to a
  pair of D2-branes with uniformly embedded D0- and anti-D0-branes.}
\label{f:crossed}
\end{figure}

The system we want to consider is a simple generalization of the (D2 +
D0)-brane configuration we described using Yang-Mills theory in
Section 2.4.  Consider a pair of D2-branes wrapped on a two-torus, one
of which has a D0-brane embedded in it as a constant positive magnetic
flux, and the other of which has an anti-D0-brane within it described
by a constant negative magnetic flux.  We take the two dimensions of
the torus to be $L_1, L_2$.  Following the discussion of Section 2.4,
this configuration is equivalent under T-duality in the $L_2$
direction to a pair of crossed D1-branes (see Figure~\ref{f:crossed}).
The Born-Infeld energy of this configuration is
\begin{eqnarray}
E_{{\rm BI}}  & = & 2 \sqrt{(\tau_2L_1 L_2)^2 + \tau_0^2}  \nonumber\\
 & = & \frac{1}{g}  \left[
\frac{2L_1 L_2}{\sqrt{2 \pi}}  + \frac{(2 \pi)^{3/2}}{L_1 L_2}  +
 \cdots \right] \label{eq:energy-bi}
\end{eqnarray}
in units where $2 \pi \alpha' = 1$.  The second term in the last line
corresponds to the Yang-Mills approximation.  In this approximation
(dropping the D2-brane energy) the energy is
\begin{equation}
E_{{\rm YM}} = \frac{\tau_2}{4}  \int {\rm Tr}\; F_{\alpha \beta}
F^{\alpha \beta} = \frac{1}{4 \sqrt{2 \pi} g}  \int {\rm Tr}\;
F_{\alpha \beta}
F^{\alpha \beta}\,.
\end{equation}

We are interested in studying this configuration in the Yang-Mills
approximation, in which we have a $U(2)$ theory on $T^2$ with field
strength
\begin{equation}
F_{12} = \left(\begin{array}{cc}
\frac{2 \pi}{ L_1 L_2}  & 0\\
0 &-\frac{2 \pi}{ L_1 L_2} 
\end{array}\right)
= \frac{2 \pi}{ L_1 L_2}  \tau_3 \,.
\end{equation}
This field strength can be realized as the curvature of a linear gauge
field
\begin{equation}
A_1 = 0, \;\;\;\;\;
A_2 =\frac{2 \pi}{ L_1 L_2}  \xi\tau_3 
\label{eq:unstable-background}
\end{equation}
which satisfies the boundary conditions
\begin{equation}
A_j (L, \xi_2) = \Omega (i \partial_j+ A_j (0, \xi_2)) \Omega^{-1}
\label{eq:bc-tachyon}
\end{equation}
where
\begin{equation}
\Omega = e^{2 \pi i ( \xi_1/L_2) \tau_3} \,.
\end{equation}

\begin{figure}
\epsfig{file=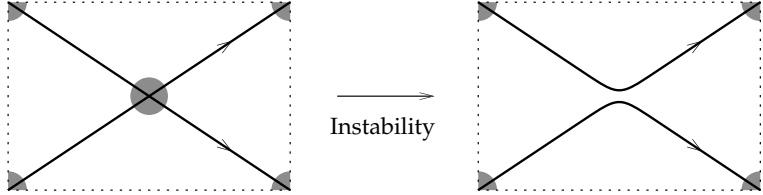,width=15cm}
\caption[x]{\footnotesize The brane-antibrane instability of a
  D0-D$\bar{0}$ system embedded in two D2-branes, as seen in the
  T-dual D1-brane picture.}
\label{f:instability}
\end{figure}

It is easy to check that this configuration indeed satisfies
\begin{equation}
E_{{\rm YM}} = \frac{1}{2g}  \frac{(2 \pi)^{3/2}}{L_1 L_2}  {\rm Tr}\;
\tau_3^2 = \frac{1}{g}  \frac{(2 \pi)^{3/2}}{L_1 L_2} 
\label{eq:unstable-energy}
\end{equation}
as desired from (\ref{eq:energy-bi}).
Since, however, 
\begin{equation}
{\rm Tr}\; F_{\alpha \beta} = 0,
\end{equation}
the gauge field we are considering is in the same topological
equivalence class as $F = 0$.  This corresponds to the fact that the
D0-brane and anti-D0-brane can annihilate.  To understand the
appearance of the tachyon, we can consider the spectrum of excitations
$\delta A_\alpha$ around the background
(\ref{eq:unstable-background}) 
 \cite{Hashimoto-Taylor}.  The
eigenvectors of the quadratic mass terms in this background are
described by theta functions on the torus satisfying boundary
conditions related to (\ref{eq:bc-tachyon}).  There are precisely two
elements in the spectrum with the negative eigenvalue $-4 \pi/L_1
L_2$.  These theta functions, given explicitly in
 \cite{Hashimoto-Taylor}, are  tachyonic modes of the theory which
are associated with the annihilation of the positive and negative
fluxes encoding the D0- and anti-D0-brane.  These tachyonic modes are
perhaps easiest to understand in the dual configuration, where they
provide a direction of instability in which the two crossed D1-branes
reconnect as in Figure~\ref{f:instability}.
In the T-dual picture it is also interesting to note that the
two tachyonic modes of the gauge field have support which is localized near
the two brane intersection points.  These modes have off-diagonal form
\begin{equation}
\delta A_t \sim \left(\begin{array}{cc}
0 & \star\\
\star & 0
\end{array} \right) \,.
\end{equation}
This form of the tachyonic modes naturally encodes our geometric
understanding of these modes as reconnecting the two D1-branes near
the intersection point.

The full Yang-Mills action around the background
(\ref{eq:unstable-background}) can be written as a quartic function of
the mass eigenstates around this background.  Written in terms of
these modes, there are nontrivial cubic and quartic terms which couple
the tachyonic modes to all the massive modes in the system.  If we
integrate out the massive modes, we know from the topological
reasoning above that an effective potential arises for the tachyonic
mode $A_t$, with a maximum value of (\ref{eq:unstable-energy}) and a
minimum value of 0.  This system is highly analogous to the bosonic
open string tachyon we will discuss in the remainder of these
lectures.  Our current understanding of the bosonic string through
bosonic string field theory is analogous to that of someone who only
knows the Yang-Mills theory around the background
(\ref{eq:unstable-background}) in terms of a complicated quartic
action for an infinite family of modes.  Without knowledge of the
topological structure of the theory, and given only a list of the
coefficients in the quartic action, such an individual would have to
systematically calculate the tachyon effective potential by explicitly
integrating out all the massive modes one by one.  This would give a
numerical approximation to the minimum of the effective potential,
which could be made arbitrarily good by raising the mass of the cutoff
at which the effective action is computed.  It may be helpful to keep
this example system in mind in the following sections, where an
analogous tachyonic system is considered in string field theory.  For
further discussion of this unstable configuration in Yang-Mills
theory, see \cite{Hashimoto-Taylor,gns}.


\section{Open string field theory and the Sen conjectures}
\label{conjectures}

The discussion of the previous sections gives us an overview of string
theory, and an example of how tachyons appear in a simple gauge theory
context, when an unstable brane-antibrane configuration is embedded in
a higher-dimensional brane.  We now turn our attention back to string
theory, where the appearance of a tachyon necessitates a
nonperturbative approach to the theory.  In subsection 4.1, we
review the BRST quantization approach to the bosonic open string.
Subsection 4.2 describes Witten's cubic open string field theory,
which gives a nonperturbative off-shell definition to the open bosonic
string.  In subsection 4.3 we describe Sen's conjectures on tachyon
condensation in the open bosonic string.


\subsection{The bosonic open string}
\label{sec:bosonic-string}

In this subsection we review the quantization of the open bosonic
string.  For further details see the textbooks by Green, Schwarz, and
Witten \cite{gsw} and by Polchinski \cite{Polchinski-string}.
The bosonic open string can be quantized using the BRST quantization
approach starting from the action
\begin{equation}
S = -\frac{1}{4 \pi \alpha'}  \int \sqrt{-\gamma} \gamma^{ab}
\partial_a X^\mu \partial_b X_\mu,
\label{eq:string-action}
\end{equation}
where $\gamma$ is an auxiliary dynamical metric on the world-sheet.
This action can be gauge-fixed to conformal gauge $\gamma_{ab} \sim
\delta_{ab}$.  Using the BRST approach to gauge fixing introduces
ghost and antighost fields $c^{\pm} (\sigma), b_{\pm \pm} (\sigma)$.  The
gauge-fixed action, including ghosts, then becomes
\begin{equation}
S = -\frac{1}{4 \pi \alpha'}  \int 
\partial_a X^\mu \partial^a X_\mu
+ \frac{1}{ \pi}  \int \left( b_{++} \partial_-c^+ + b_{--} \partial_+
c^-\right)\,. 
\label{eq:gauge-fixed-action}
\end{equation}

The matter fields $X^\mu$ can be expanded in modes using
\begin{equation}
X^\mu (\sigma, \tau) = x_0^\mu + l_s^2 p^\mu \tau +
\sum_{n \neq 0}
\frac{i l_s}{n}  \alpha^\mu_n \cos (n \sigma) e^{-in \tau}\,.
\label{eq:mode-decomposition}
\end{equation}
Throughout the remainder of these notes we will use the convention
\begin{equation}
\alpha' = \frac{l_s^2}{ 2}  = 1\,,
\end{equation}
so that $l_s = \sqrt{2}$.
In the quantum theory, $x_0^\mu$ and $p^\mu$ obey the canonical
commutation relations 
\begin{equation}
[x^\mu_0, p^\nu] = i \eta^{\mu \nu}\,.
\end{equation}
The $\alpha^\mu_n$'s with negative/positive
values of $n$ become raising/lowering operators for the oscillator
modes on the string, and satisfy the
commutation relations
\begin{equation}
[\alpha^\mu_m, \alpha^\nu_n] = m \eta^{\mu \nu} \delta_{m + n, 0}\,.
\end{equation}
We will often use the canonically normalized raising and lowering
operators
\begin{equation}
a^\mu_n = \frac{1}{ \sqrt{|n |}}  \alpha^{\mu}_n
\end{equation}
which obey the commutation relations
\begin{equation}
[a^\mu_m, a^\nu_n] =  \eta^{\mu \nu} \delta_{m + n, 0}\,.
\end{equation}
The raising and lowering operators satisfy $(\alpha^\mu_n)^{\dagger} =
\alpha^\mu_{-n}, (a^\mu_n)^{\dagger} =
a^\mu_{-n}$.
We will also frequently use position modes $x_n$ for $n \neq 0$ and
raising and lowering operators $a_0, a^{\dagger}_0$ for the zero
modes.  These are related to the modes in
(\ref{eq:mode-decomposition}) through (dropping space-time indices)
\begin{eqnarray}
x_n & = &  \frac{i}{ \sqrt{n}} (a_n- a^{\dagger}_n) \\
x_0 & = &  \frac{i}{\sqrt{2}}  (a_0-a_0^{\dagger})
\nonumber
\end{eqnarray}

The ghost and antighost fields can be decomposed into modes through
\begin{eqnarray}
c^{\pm} (\sigma, \tau) & = &  \sum_{n}c_n e^{\mp in (\sigma \pm \tau)} \\
b_{\pm \pm} (\sigma, \tau) & = &  \sum_{n}b_n e^{\mp in (\sigma \pm \tau)} \,.
\nonumber
\end{eqnarray}
The ghost and antighost modes satisfy the anticommutation relations
\begin{eqnarray}
\{c_n, b_m\} & = &  \delta_{n + m, 0}\\
\{c_n, c_m\}  =\{b_n, b_m\}  & = &  0\,. \nonumber
\end{eqnarray}

A general state in the open string Fock space can be written in the
form
\begin{equation}
\alpha^{\mu_1}_{-n_1} \cdots \alpha^{\mu_i}_{-n_i} \;
c_{-m_1} \cdots c_{-m_j} \;
b_{-p_1} \cdots b_{-p_l} \; | 0; k \rangle
\end{equation}
where $| 0; k \rangle$ is the SL(2,R) invariant vacuum annihilated by
\begin{eqnarray}
b_n | 0; k \rangle  & = &  0, \; \; \; n \geq -1\\
c_n | 0; k \rangle  & = &  0, \; \; \; n \geq 2\\
\alpha^\mu_{-n} | 0; k \rangle  & = &  0, \; \; \; n \geq 1
\end{eqnarray}
with momentum 
\begin{equation}
p^\mu | 0; k \rangle = k^\mu | 0; k \rangle\,.
\end{equation}
We will often write the zero momentum vacuum $| 0; k = 0 \rangle$
simply as $| 0 \rangle$.  This vacuum is taken by convention to have
ghost number 0, and satisfies
\begin{equation}
\langle 0; k | c_{-1} c_0c_1 | 0 \rangle =  \delta (k)
\end{equation}
For string field theory we will also find it
convenient to work with the vacua of ghost number 1 and 2
\begin{eqnarray}
G = 1: & \hspace*{0.1in} &  | 0_1  \rangle = c_1| 0 \rangle\\
G = 2: & \hspace*{0.1in} &  | 0_2  \rangle = c_0 c_1| 0 \rangle\,.
\end{eqnarray}
In the notation of Polchinski \cite{Polchinski-string}, these two vacua
are written as
\begin{eqnarray}
| 0_1 \rangle & = & | 0 \rangle_m \otimes |\!\downarrow \rangle \\
| 0_2 \rangle & = & | 0 \rangle_m \otimes |\!\uparrow \rangle  \nonumber
\end{eqnarray}
where $| 0 \rangle_m$ is the matter vacuum and $|\!\downarrow\rangle,
|\!\uparrow\rangle$ are the ghost vacua annihilated by $b_0, c_0$.

The BRST operator of this theory is given by
\begin{equation}
Q_B = \sum_{n = -\infty}^{\infty}  c_nL_{-n}^{({\rm m})}
+ \sum_{n, m = -\infty}^{ \infty}  \frac{(m-n)}{2} 
:c_mc_nb_{-m-n}:-c_0
\label{eq:BRST}
\end{equation}
where the matter Virasoro operators are given by
\begin{equation}
L_q^{({\rm m})} = \left\{
\begin{array}{ll}
\frac{1}{2}\sum_{n}\alpha^\mu_{q-n} \alpha_{\mu \; n}, & q \neq 0\\
p^2 + \sum_{n = 1}^{ \infty}  \alpha^\mu_{-n} \alpha_{\mu
\; n}, \;\;\;\;\;& q = 0
\end{array}
\right.
\end{equation}

Some useful features of the BRST operator $Q =Q_B$ include:
\begin{itemize}
\item $Q^2 = 0$; {\it i.e.}, the BRST operator is nilpotent.  This
  identity relies on a cancellation between matter and ghost terms
  which only works in dimension $D = 26$ for the bosonic theory.
\item $\{Q, b_0\} = L_0^{({\rm m})} + L_0^{({\rm g})} -1$.
\item $Q$ has ghost number 1, so acting on a state $| s \rangle$ of
  ghost number $G$ gives a state $Q | s \rangle$ of ghost number $G
  + 1$.
\item The physical states of the theory are given by the cohomology of
  $Q$ at ghost number 1
\begin{eqnarray}
{\cal H}_{{\rm phys}} & = & {\cal H}_{{\rm closed}}/{\cal H}_{{\rm
    exact}}
\nonumber\\
 & = & \{| \psi \rangle:Q | \psi \rangle = 0\}/
\left( | \psi \rangle \sim | \psi \rangle + Q | \chi \rangle \right)
\end{eqnarray}
\item Physical states can be chosen as representatives of each
  cohomology class so that they are all annihilated by $b_0$.
\end{itemize}

It is often convenient to separate out the ghost zero-modes, writing
$Q = c_0 L_0+ b_0M + \tilde{Q}$, where (momentarily reinstating $ \alpha'$)
\begin{equation}
L_0 = \sum_{n = 1}^{ \infty} 
\left( \alpha_{-n} \alpha_n + nc_{-n} b_n + nb_{-n} c_n \right) +
\alpha' p^2 -1
\end{equation}
In this expression the term in parentheses is simply the oscillator
number operator, measuring the level of a given state.

Some simple examples of physical states include the tachyon state
\begin{equation}
| 0_1; p \rangle
\end{equation}
which is physical when $p^2 = 1/\alpha' = -M^2$,
and the massless gauge boson
\begin{equation}
\epsilon_\mu \alpha^\mu_{- 1}| 0_1; p \rangle
\label{eq:gauge-boson}
\end{equation}
which is physical when $p^2 = M^2 = 0$, for transverse polarizations
$p \cdot \epsilon = 0$.  Note that the transverse polarization
condition follows from the appearance of a term proportional to
$c_{-1} p \cdot \alpha_1$ in $\tilde{Q}$, which must annihilate the
state (\ref{eq:gauge-boson})


\subsection{Witten's cubic bosonic SFT}
\label{sec:Witten-SFT}

The discussion of the previous subsection leads to a systematic
quantization of the open bosonic string in the conformal field theory
framework.  Using this approach it is possible, in principle, to
calculate an arbitrary perturbative on-shell scattering amplitude for
physical string states.  To study tachyon condensation in string
theory, however, we require a nonperturbative, off-shell formalism for
the theory--- a string field theory.

A very simple form for the
off-shell open bosonic string field theory action was proposed by
Witten in 1986 \cite{Witten-SFT}
\begin{equation}
S = -\frac{1}{2}\int \Psi \star Q \Psi -\frac{g}{3}  \int \Psi \star
\Psi \star \Psi\,.
\label{eq:SFT-action}
\end{equation}
This action has the general form of a Chern-Simons theory on a
3-manifold, although for string field theory there is no explicit
interpretation of the integration in terms of a concrete 3-manifold.
In Eq.~(\ref{eq:SFT-action}), $g$ is interpreted as the string
coupling constant.  The field $\Psi$ is a string field, which takes
values in a graded algebra ${\cal A}$.  Associated with the algebra
${\cal A}$ there is a star product
\begin{equation}
\star:{\cal A} \otimes{\cal A} \rightarrow{\cal A}, \;\;\;\;\;
\end{equation}
under which the degree $G$ is additive ($G_{\Psi \star \Phi} = G_\Psi
+ G_\Phi$).  There is also a BRST operator
\begin{equation}
Q:{\cal A} \rightarrow{\cal A}, \;\;\;\;\;
\end{equation}
of degree one ($G_{Q \Psi} = 1 + G_\Psi$).  String fields can be
integrated using
\begin{equation}
\int:{\cal A} \rightarrow {\bb C}\,.
\end{equation}
This integral vanishes for all $\Psi$ with degree $G_\Psi \neq 3$.

The elements $Q, \star, \int$ defining the string field theory are
assumed to satisfy the following axioms:
\vspace*{0.15in}

\noindent {\bf (a)} Nilpotency of $Q$: $\;Q^2 \Psi = 0, \;\; \; \forall \Psi
\in{\cal A}$.
\vspace*{0.08in}

\noindent {\bf (b)} $\int Q\Psi = 0, \; \; \; \forall \Psi \in{\cal A}$.
\vspace*{0.08in}

\noindent {\bf (c)} Derivation property of $Q$:\\
\hspace*{0.4in}$\;Q (\Psi \star \Phi) = (Q \Psi) \star \Phi +
(-1)^{G_\Psi} \Psi \star (Q \Phi), \; \; \forall \Psi, \Phi \in{\cal A}$.
\vspace*{0.08in}

\noindent {\bf (d)} Cyclicity:  $\;\int \Psi \star \Phi = (-1)^{G_\Psi
G_\Phi} \int \Phi \star \Psi, \; \; \; \forall \Psi, \Phi \in{\cal A}$.
\vspace*{0.08in}

\noindent {\bf (e)}  Associativity:  $(\Phi \star \Psi) \star \Xi =
\Phi \star (\Psi \star \Xi), \; \; \;
\forall \Phi, \Psi, \Xi \in{\cal A}$.
\vspace*{0.15in}

When these axioms are satisfied, the action (\ref{eq:SFT-action}) is
invariant under the gauge transformations
\begin{equation}
\delta \Psi = Q \Lambda + \Psi\star \Lambda - \Lambda \star \Psi
 \label{eq:SFT-gauge}
\end{equation}
for any gauge parameter $\Lambda \in{\cal A}$ with ghost number 0.

When the string coupling $g$ is taken to vanish, the equation of
motion for the theory defined by (\ref{eq:SFT-action}) simply becomes
$Q \Psi = 0$, and the gauge transformations (\ref{eq:SFT-gauge})
simply become 
\begin{equation}
\delta \Psi = Q \Lambda\,.
\end{equation}
Thus, when $g = 0$ this string field theory gives precisely the structure
needed to describe the free bosonic string.  The motivation for
introducing the extra structure in (\ref{eq:SFT-action}) was to
find a simple interacting extension of the free theory, consistent
with the perturbative  expansion of open bosonic string theory.

Witten presented this formal structure and argued
that all the needed axioms are satisfied when ${\cal A}$ is taken to
be the space of string fields
\begin{equation}
{\cal A} =\{\Psi[x (\sigma); c (\sigma), b
(\sigma)]\}
 \label{eq:string-functionals}
\end{equation}
which can be described as functionals of the matter, ghost and
antighost fields describing an open string in 26 dimensions with $0
\leq \sigma \leq \pi$.  Such a string field can be written as a formal
sum over open string Fock space states with coefficients given by an
infinite family of space-time fields
\begin{equation}
\Psi =
\int d^{26}p \;
\left[ \phi (p)\; | 0_1; p \rangle + A_\mu (p) \; \alpha^\mu_{-1} | 0_1; p
\rangle + \cdots \right]
 \label{eq:field-expansion}
\end{equation}
Each Fock space state is associated with a given string functional,
just as the states of a harmonic oscillator are associated with
wavefunctions of a particle in one dimension.  For example, the matter
ground state $| 0 \rangle_m$ annihilated by $a_n$ for all $n \geq 1$
is associated (up to a constant $C$) with the functional of matter
modes
\begin{equation}
| 0 \rangle_m \rightarrow
C \exp \left( -\frac{1}{ 4}\sum_{n > 0}^{ \infty} 
nx_n^2 \right)\,.
\end{equation}

For Witten's cubic string field theory, the BRST operator $Q$ in
(\ref{eq:SFT-action}) is the usual open string BRST operator $Q_B$,
given in (\ref{eq:BRST}).  The star product $\star$ acts on a pair
of functionals $\Psi, \Phi$ by gluing the right half of one string to
the left half of the other using a delta function interaction

\begin{center}
\begin{picture}(100,60)(- 50,- 30)
\put(-40,20){\line(1,0){37}}
\put(40,20){\line(-1,0){37}}
\put(-3,20){\line( 0, -1){ 37}}
\put(3,20){\line( 0, -1){ 37}}
\put(-20,2){\makebox(0,0){$\Psi$}}
\put(20, 2){\makebox(0,0){$\Phi$}}
\put(0,-25){\makebox(0,0){$\delta$}}
\end{picture}
\end{center}

This star product factorizes into separate matter and ghost parts.  In
the matter sector, the star product is given by the formal functional
integral
\begin{eqnarray}
\lefteqn{\left(\Psi \star   \Phi\right) [z(\sigma)]} \label{eq:mult} \\ & 
\equiv&
\int
\prod_{{0} \leq \tilde{\tau} \leq {\pi\over 2}} dy(\tilde{\tau}) \; dx
(\pi -\tilde{\tau}) 
\prod_{{\pi\over 2} \leq
\tau \leq \pi}
\delta[x(\tau)-y(\pi-\tau)]
\;   \Psi [x(\tau)]  \Phi [y(\tau)]\, ,\nonumber\\
& &
\hspace*{1.2in}x(\tau)  = z(\tau) \quad {\rm for} \quad {0} \leq \tau \leq
{\pi\over 2}\, , 
\nonumber\\
& &
\hspace*{1.2in}y(\tau)  = z(\tau)\quad {\rm for} \quad   {\pi\over 2} \leq
\tau \leq \pi\, . 
\nonumber
\end{eqnarray}
Similarly, the integral over a string field factorizes into matter and
ghost parts, and in the matter sector is given by
\begin{equation}
\int \Psi = \int \prod_{0 \leq \sigma \leq \pi} dx (\sigma) \;
\prod_{0 \leq
\tau \leq \frac{\pi}{2} }
\delta[x(\tau)-x(\pi-\tau)] \;\Psi[x (\tau)]\,.
\label{eq:integral-p}
\end{equation}
This corresponds to gluing the left and right halves  of the string
together with a delta function interaction

\begin{center}
\begin{picture}(100,60)(- 50,- 30)
\put(-3,20){\line(1,0){6}}
\put(-3,20){\line( 0, -1){ 37}}
\put(3,20){\line( 0, -1){ 37}}
\put(0,-25){\makebox(0,0){$\delta$}}
\put(0,27){\makebox(0,0){$\Psi$}}
\end{picture}
\end{center}

The ghost sector of the theory is defined in a similar fashion, but
has an anomaly due to the curvature of the Riemann surface describing
the three-string vertex.  The ghost sector can be described either in
terms of fermionic ghost fields $c (\sigma), b (\sigma)$ or through
bosonization in terms of a single bosonic scalar field $\phi_g
(\sigma)$.  From the functional point of view of Eqs.~(\ref{eq:mult},
\ref{eq:integral-p}), it is easiest to describe the ghost sector in the
bosonized language.  In this language, the ghost fields $b (\sigma)$
and $c (\sigma)$ are replaced by the scalar field $\phi_g (\sigma)$, and
the star product in the ghost
sector is given by (\ref{eq:mult}) with an extra insertion of $\exp
(3i \phi_g (\pi/2)/2)$ inside the integral.  Similarly, the integration
of a string field in the ghost sector is given by (\ref{eq:integral-p})
with an insertion of $\exp (-3i \phi_g (\pi/2)/2)$ inside the integral.
Witten first described the cubic string field theory using
bosonized ghosts.  While this approach is useful for some purposes, we
will use fermionic ghost fields in the remainder of these lecture notes.

The expressions (\ref{eq:mult}, \ref{eq:integral-p}) may seem rather
formal, as they are written in terms of
functional integrals.  These
expressions, however, can be given precise meaning when described in
terms of creation and annihilation operators acting on the string Fock
space.  
In the Fock space language, the integral of a star product of two or
three fields is described in terms of two- and three-string vertices
\begin{equation}
\langle V_2 |\in {\cal H}^* \otimes {\cal H}^*, \;\;\;\;\;
\langle V_3 | \in \left({\cal H}^* \right)^3
\label{eq:23-vertices}
\end{equation}
so that
\begin{eqnarray}
\int \Phi \star \Psi & \rightarrow &  \langle V_2 |
\left( |\Phi \rangle \otimes | \Psi \rangle \right)\\
\int \Psi_1 \star \Psi_2 \star \Psi_3 & \rightarrow &  \langle V_3 |
\left( |\Psi_1 \rangle \otimes |\Psi_2 \rangle \otimes | \Psi_3
\rangle \right) \nonumber
\end{eqnarray}
In the next section we will give explicit forms for the two- and
three-string vertices (\ref{eq:23-vertices}).  In terms of these
vertices, the string field theory action becomes
\begin{equation}
S = -\frac{1}{2}\langle V_2 | \Psi, Q \Psi \rangle
-\frac{g}{3}   \langle V_3 | \Psi, \Psi, \Psi \rangle\,.
\label{eq:action-Fock}
\end{equation}
This action is often written using the BPZ dual $\langle \Psi |$ of
the string field $| \Psi \rangle$, defined by the conformal map $z
\rightarrow -1/z$, as
\begin{equation}
S = -\frac{1}{2}\langle \Psi | Q  \Psi \rangle
-\frac{g}{3}  \langle \Psi | \Psi \star \Psi \rangle\,.
\end{equation}
In the remainder of these lectures, however, we will use the form
(\ref{eq:action-Fock}).
Using explicit formulae for the vertices (\ref{eq:23-vertices}) and
the string field expansion (\ref{eq:field-expansion}) leads to the
full string field theory action, given by  an
off-shell action in the target space-time for an infinite family of
fields $\phi (p), A_\mu (p), \ldots$
We discuss this action in more detail in Section 5.


\subsection{The Sen conjectures}
\label{sec:conjectures}

The existence of the tachyonic mode in the open bosonic string
indicates that the standard choice of perturbative vacuum for this
theory is unstable.  In the early days of the subject, there was some
suggestion that this tachyon could condense, leading to a more stable
vacuum (see for example \cite{Bardakci-tachyon}).  Kostelecky and
Samuel argued early on that the stable vacuum could be identified in
string field theory in a systematic way \cite{ks-open}, however there
was no clear physical picture for the significance of this stable vacuum.
In 1999, Ashoke Sen reconsidered the problem of tachyons in string
field theory.  Sen suggested that the open bosonic string should
really be thought of as living on a D25-brane, and hence that the
perturbative vacuum for this string theory should have a nonzero
vacuum energy associated with the tension of this D25-brane.  He
suggested that the tachyon is simply the instability mode of the
D25-brane, which carries no conserved charge and hence is not expected
to be stable, as discussed in section 3.  Sen furthermore suggested
that Witten's cubic open string field theory is a natural framework to
use to study this tachyon, and that this string field theory should
give an analytic description of the true vacuum.  More precisely, Sen
made the following 3 conjectures \cite{Sen-universality}:
\begin{enumerate}
\item Witten's classical open string field theory should have a
  locally stable nontrivial vacuum solution.  The energy density of
  this vacuum should be given by  the D25-brane tension
\begin{equation}
\frac{\Delta E}{V}  =T_{25} = -\frac{1}{2 \pi^2 g^2} \,.
\end{equation}
\item Lower-dimensional D-branes should exist as solitonic solutions
  of SFT which break part of the Lorentz symmetry of the perturbative
  vacuum.
\item Open strings should decouple from the theory in the nontrivial
  vacuum, since the D25-brane is absent in this vacuum.
\end{enumerate}

In Section 6 of these lectures we discuss the evidence for these
conjectures, focusing particularly on the first and third
conjectures.  First, however, we need to develop the technical tools
to do specific calculations in string field theory.


\section{Basics of SFT}
\label{sec:overlaps}

In this section, we give a more detailed discussion of Witten's open
bosonic string field theory.  Subsection 5.1 is a warmup, in which we
review some basic features of the simple harmonic oscillator and
discuss squeezed states.  In Subsection 5.2 we derive the two-string
vertex, and in subsection 5.3 we give an explicit formula for the
three-string vertex.  In subsection 5.4 we put these pieces together
and discuss the calculation of the full SFT action.  5.5 contains a
brief description of some more general features of Witten's open
bosonic string field theory.
For more details about this string field theory, the reader is
referred to the reviews \cite{lpp,Thorn,Gaberdiel-Zwiebach}.


\subsection{Squeezed states and the simple harmonic oscillator}
\label{sec:warmup}

Let us consider a simple harmonic oscillator with annihilation
operator
\begin{equation}
a = -i \left( \sqrt{\frac{\alpha}{ 2}}x + \frac{1}{ \sqrt{2 \alpha}}
  \partial_x   \right)
\end{equation}
and ground state
\begin{equation}
| 0 \rangle = \left( \frac{\alpha}{ \pi}  \right)^{1/4} e^{-\alpha x^2/2}\,.
\end{equation}
In the harmonic oscillator basis $|n \rangle$, the Dirac position basis
states $| x \rangle$ have a squeezed state form
\begin{equation}
| x \rangle = \left( \frac{\alpha}{ \pi}  \right)^{1/4}
\exp\left(-\frac{\alpha}{ 2}x^2
-i \sqrt{2 \alpha} a^{\dagger} x +\frac{1}{2} (a^{\dagger})^2 \right)
| 0 \rangle\,. 
\end{equation}

A general wavefunction is associated with a state through the
correspondence
\begin{equation}
f (x) \rightarrow \int_{-\infty}^\infty dx \;f (x) | x \rangle \,.
\label{eq:function-state}
\end{equation}
In particular, we have
\begin{eqnarray}
\delta (x) & \rightarrow & 
\left( \frac{\alpha}{ \pi}  \right)^{1/4}
\exp\left(\frac{1}{2} (a^{\dagger})^2 \right)
| 0 \rangle \label{eq:squeezed-d1}
\\
1 & \rightarrow &  \int dx \; | x \rangle =
\left( \frac{4\pi}{ \alpha}  \right)^{1/4}
\exp\left(-\frac{1}{2} (a^{\dagger})^2 \right)
| 0 \rangle\nonumber
\end{eqnarray}
This shows that the delta and constant functions both have squeezed
state representations in terms of the harmonic oscillator basis.
The norm of a squeezed state
\begin{equation}
| s \rangle =\exp\left(\frac{1}{2} s (a^{\dagger})^2 \right)
| 0 \rangle
\end{equation}
is given by
\begin{equation}
\langle s| s \rangle =
\frac{1}{ \sqrt{1-s^2}} 
\end{equation}
Thus, the states (\ref{eq:squeezed-d1}) are non-normalizable (as we
would expect), however they are right on the border of
normalizability.  As for the Dirac basis states $| x \rangle$, which
are computationally useful although
technically not well-defined states in the single-particle Hilbert
space, we expect that many calculations using the states
(\ref{eq:squeezed-d1}) will give sensible physical answers.

It will be useful for us to generalize the foregoing considerations in
several ways.  A particularly simple generalization arises when we
consider a pair of degrees of freedom $x, y$ described by a two-harmonic
oscillator Fock space basis.  In such a basis, repeating the preceding
analysis leads us to a function-state correspondence for the delta
functions relating $x, y$ of the form
\begin{equation}
\delta (x \pm y) \rightarrow
\exp\left(\pm\frac{1}{2} a^{\dagger}_{(x)}a^{\dagger}_{(y)}\right)
\left( | 0 \rangle_x \otimes | 0 \rangle_y \right)\,.
\label{eq:squeezed-two}
\end{equation}
we will find these squeezed state expressions very useful in
describing the two- and three-string vertices of Witten's open string
field theory.


\subsection{The two-string vertex $| V_2 \rangle$}
\label{sec:v2}

We can immediately apply the oscillator formulae from the preceding
section to calculate the two-string vertex.  Recall that the matter
fields are expanded in modes through
\begin{equation}
x (\sigma) = x_0 + \sqrt{2} \sum_{n = 1}^{ \infty}  x_n \cos n \sigma\,.
\end{equation}
(We suppress Lorentz indices in most of this section for clarity.)
Using this mode decomposition, we associate the string field
functional $\Psi[x (\sigma)]$ with a function $\Psi (\{x_n\})$ of the
infinite family of string oscillator mode  amplitudes.  The overlap
integral combining (\ref{eq:integral-p}) and (\ref{eq:mult}) can then
be expressed in modes as
\begin{equation}
\int \Psi \star \Phi = \int \prod_{n = 0}^{ \infty}  dx_ndy_n
\; \delta (x_n-(-1)^ny_n) \Psi (\{x_n\}) \Phi (\{y_n\})\,.
\end{equation}
Geometrically this just encodes the overlap condition $x (\sigma) = y
(\pi -\sigma)$ described through

\begin{center}
\centering
\begin{picture}(100,40)(- 50,- 20)
\put( 20,2){\vector( -1,0){40}}
\put( -20,-2){\vector( 1,0){40}}
\put(0,-10){\makebox(0,0){$\Psi$}}
\put(0, 10){\makebox(0,0){$\Phi$}}
\end{picture}
\end{center}

From (\ref{eq:squeezed-two}), it follows that we can write the
two-string vertex as a squeezed state
\begin{equation}
\langle V_2 |_{{\rm matter}} =
\left(\langle 0 | \otimes \langle 0 |\right)
\exp\left( \sum_{n, m = 0}^{ \infty} 
-a_n^{(1)}  C_{nm} a_m^{ (2)} \right)
\label{eq:v2o}
\end{equation}
where $C_{nm} = \delta_{nm} (-1)^n$ is an infinite-size matrix
connecting the oscillator modes of the two single-string Fock spaces,
and the sum is taken over all oscillator modes including zero.  In the
expression (\ref{eq:v2o}), we have used the formalism in which $| 0
\rangle$ is the vacuum annihilated by $a_0$.  To translate this
expression into a momentum basis, we use only $n, m > 0$, and replace
\begin{equation}
\left(\langle 0 | \otimes \langle 0 |\right)
\exp\left(-a^{(1)}_0a^{(2)}_0\right) \rightarrow
\int d^{26} p
\left( \langle 0; p | \otimes \langle 0; -p | \right)\,.
\end{equation}

The extension of this analysis to ghosts is straightforward.  For the
ghost and antighost respectively, the overlap conditions corresponding
with
$x_1 (\sigma) = x_2 (\pi -\sigma)$ are \cite{Gross-Jevicki-12}
$c_1 (\sigma) = -c_2 (\pi -\sigma)$ and
$b_1 (\sigma) = b_2 (\pi -\sigma)$.  This leads to the overall formula
for the two-string vertex
\begin{equation}
\langle V_2 | = \int d^{26} p
\left( \langle 0; p | \otimes \langle 0; -p | \right)
(c^{(1)}_0 + c^{(2)}_0)\exp\left(
-\sum_{n = 1}^{ \infty} (-1)^n
[a^{(1)}_n a^{(2)}_n+c^{(1)}_n b^{(2)}_n+c^{(2)}_n b^{(1)}_n] \right)\,.
 \label{eq:v2}
\end{equation}
This expression for the two-string vertex can also be derived directly
from the conformal field theory approach, computing the two-point
function of an arbitrary pair of states on the disk.


\subsection{The three-string vertex $| V_3 \rangle$}
\label{sec:v3}

The three-string vertex, which is associated with the three-string
overlap diagram
\begin{center}
\centering
\begin{picture}(100,60)(- 50,- 30)
\put( 20,14){\line( -2,-1){20}}
\put(0, 4){\vector( -2,1){20}}
\put(-20,9){\line( 2, -1){18}}
\put( -2, 0 ){\vector( 0, -1){ 20}}
\put( 2,-20){\line( 0, 1){ 20}}
\put(2,0){\vector( 2,1){18}}
\put(-15,-10){\makebox(0,0){$\Psi_2$}}
\put(0, 15){\makebox(0,0){$\Psi_1$}}
\put(15, -10){\makebox(0,0){$\Psi_3$}}
\end{picture}
\end{center}
can be computed in a very similar fashion to the two-string vertex
above.  The details of the calculation, however, are significantly
more complicated.  There are several different ways to carry out the
calculation.  One approach is to first rewrite the modes $\cos n
\sigma$ on the full string in terms of modes $l, r$ on the two halves
of the string with $\sigma < \pi/2, \sigma > \pi/2$.  This rewriting
can be accomplished using an infinite orthogonal transformation matrix
$X$.  The delta function overlap condition can then be applied to the
half-string modes as above, giving a squeezed state expression for $|
V_3 \rangle$ with a squeezing matrix which can be expressed in terms
of $X$.
The
three-string vertex can also be computed using the conformal
field theory approach.  
The three-string vertex was computed using various versions of these
approaches in \cite{Gross-Jevicki-12,cst,Samuel,Ohta,Shelton}
\footnote{Another interesting approach to  understanding the cubic
  vertex has been explored extensively since these lectures were
  given.  By diagonalizing the Neumann matrices, the star product
  encoded in the 3-string vertex takes the form of a continuous Moyal
  product.  This simplifies the complexity of the cubic vertex, but at
  the cost of complicating the propagator.  For a recent review of
  this work and further references, see \cite{Bars}}.

In these lectures we will not have time to go through a detailed
derivation of the three-string vertex using any of these
methods\footnote{A more detailed discussion of the derivation of the
Neumann coefficients using CFT and oscillator methods will appear in
\cite{Taylor-Zwiebach}}.  We simply quote the final result from
\cite{Gross-Jevicki-12,RSZ-2}.  Like the two-string vertex, the
three-string vertex takes the  form of a squeezed state
\begin{eqnarray}
\langle V_3 | & = & \int d^{26} p^{(1)}d^{26} p^{(2)}d^{26} p^{(3)}\;
\left( \langle 0; p^{(1)} | \otimes
\langle 0; p^{(2)} | \otimes \langle 0; p^{(3)} | \right)
\delta (p^{(1)} + p^{(2)} + p^{(3)}) c_0^{(1)}c_0^{(2)}c_0^{(3)}
\nonumber\\
& &
\kappa
\exp\left(
-\frac{1}{2}\sum_{r, s = 1}^{3}
[a^{(r)}_m V^{rs}_{mn} a^{(s)}_n+
2 a^{(r)}_m V^{rs}_{m0} p^{(s)}+
p^{(r)} V^{rs}_{00}  p^{(s)}+
c^{(r)}_m X^{rs}_{mn} b^{(s)}_n] \right)\,, \label{eq:v3}
\end{eqnarray}
where $\kappa = 3^{9/2}/2^6$, and where the Neumann coefficients
$V^{rs}_{mn}, X^{rs}_{mn}$ are calculable constants given as
follows\footnote{Note that in some references, signs and various
factors in $\kappa$ and the Neumann coefficients may be slightly
different.  In some papers, the cubic term in the action is taken to
have an overall factor of $g/6$ instead of $g/3$; this choice of
normalization gives a 3-tachyon amplitude of $g$ instead of $2g$, and
gives a different value for $\kappa$.  Often, the sign in the
exponential of (\ref{eq:v3}) is taken to be positive, which changes
the signs of the coefficients $V^{rs}_{nm}, X^{rs}_{nm}$.  When the
matter Neumann coefficients are defined with respect to the oscillator
modes $\alpha_n$ rather than $a_n$, the matter Neumann coefficients
$V^{rs}_{nm}, V^{rs}_{n0}$ must be divided by $\sqrt{nm}$ and
$\sqrt{n}$.  Finally, when $\alpha'$ is taken to be $1/2$, an extra
factor of $1/\sqrt{2}$ appears for each $0$ subscript in the matter
Neumann coefficients.  }.  Define $A_n, B_n$ for $n \geq 0$ through
\begin{eqnarray}
\left( \frac{1 + ix}{1-ix} \right)^{1/3}  & = & 
\sum_{n\, {\rm even}} A_n x^n + i
\sum_{m\, {\rm odd}} A_m x^m  \label{eq:ab}\\
\left( \frac{1 + ix}{1-ix} \right)^{2/3}  & = & 
\sum_{n\, {\rm even}} B_n x^n + i
\sum_{m\, {\rm odd}} B_m x^m   \nonumber
\end{eqnarray}
These coefficients can be used to define 6-string Neumann coefficients
  $N^{r, \pm s}_{nm}$ through
\begin{eqnarray}
N^{r, \pm r}_{nm} & = & 
\left\{\begin{array}{l}
\frac{1}{3 (n \pm m)}  (-1)^n (A_nB_m \pm B_nA_m), \;\;\;\;\;
m + n\, {\rm even}, \;m \neq n\\
0, \;\;\;\;\; m + n\, {\rm odd}
\end{array} \right.\label{eq:n6}\\
N^{r, \pm (r + \sigma)}_{nm} & = & 
\left\{\begin{array}{l}
\frac{1}{6 (n \pm \sigma m)}  (-1)^{n + 1} (A_nB_m \pm \sigma B_nA_m),
\;\;\;\;\; 
m + n\, {\rm even}, \;m \neq n\\
\sigma
\frac{\sqrt{3}}{6 (n \pm \sigma m)} (A_nB_m \mp \sigma B_nA_m), \;\;\;\;\;
m + n\, {\rm odd}
\end{array}\right].\nonumber
\end{eqnarray}
where in $N^{r, \pm (r + \sigma)}$, $\sigma = \pm 1$, and $r +\sigma$
is taken modulo 3 to be between 1 and 3.  The 3-string matter Neumann
coefficients $V^{rs}_{nm}$ are then given by
\begin{eqnarray}
V^{rs}_{nm} & = &  -\sqrt{mn} (N^{r, s}_{nm} + N^{r, -s}_{nm}),
\;\;\;\;\; m \neq n,\, {\rm and}\, m, n \neq 0 \nonumber\\
V^{rr}_{nn} & = &  -\frac{1}{3}  \left[
2 \sum_{k = 0}^{n}  (-1)^{n-k} A_k^2-(-1)^n-A_n^2 \right], \;\;\;\;\;
n \neq 0 \nonumber\\
V^{r, r + \sigma}_{nn} & = &\frac{1}{2} \left[ (-1)^n-V^{rr}_{nn}
  \right], \;\;\;\;\;  n \neq 0 \label{eq:n3}\\
V^{rs}_{0n}& = & -\sqrt{2n} \left( N^{r, s}_{0n} + N^{r, -s}_{0n}
\right), \;\;\;\;\; n \neq 0\nonumber\\
V^{rr}_{00} & = & \ln (27/16) \nonumber
\end{eqnarray}
The ghost Neumann coefficients $X^{rs}_{m n}, m\geq 0, n >0$ are
given by
\begin{eqnarray}
X^{rr}_{mn} & = &  \left( -N^{r, r}_{nm} + N^{ r, -r}_{nm} \right),
 \;\;\;\;\; n \neq
 m\nonumber\\
X^{r (r \pm 1)}_{mn} & = &  m \left(\pm N^{r, r \mp 1}_{nm} \mp N^{ r, - (r
 \mp 1)}_{nm} \right), \;\;\;\;\; n \neq
 m \label{eq:x3}\\
X^{rr}_{nn} & = &  \frac{1}{3} 
\left[ -(-1)^n-A_n^2 + 2 \sum_{k = 0}^{n}  (-1)^{n-k} A_k^2 -2
 (-1)^nA_nB_n \right] \nonumber\\
X^{r (r \pm 1)}_{nn} & = &  
-\frac{1}{2}(-1)^n -\frac{1}{2} X^{rr}_{nn}
\nonumber
\end{eqnarray}

The Neumann coefficients have a number of simple symmetries.  There is
a cyclic symmetry under $r \rightarrow r + 1, s \rightarrow s + 1$,
which corresponds to the obvious geometric symmetry of rotating the
vertex.  The coefficients are also symmetric under the exchange $r
\leftrightarrow s, n \leftrightarrow m$.  Finally, there is a
``twist'' symmetry,
\begin{eqnarray}
V^{rs}_{nm} & = &  (-1)^{n + m}V^{sr}_{nm}\\
X^{rs}_{nm} & = &  (-1)^{n + m}X^{sr}_{nm}\,. \nonumber
\end{eqnarray}
This symmetry follows from the invariance of the 3-vertex
under reflection.


\subsection{Calculating the SFT action}
\label{sec:calculating}

Given the action (\ref{eq:action-Fock}) and the explicit formulae
(\ref{eq:v2}, \ref{eq:v3}) for the two- and three-string vertices, we
can in principle calculate the string field action term by term for
each of the fields in the string field expansion
\begin{equation}
\Psi =
\int d^{26}p \;
\left[ \phi (p)\; | 0_1; p \rangle + A_\mu (p) \; \alpha^\mu_{-1} | 0_1; p
\rangle + \chi ( p) b_{-1} c_0| 0_1; p \rangle + B_{\mu \nu} ( p)
\alpha^\mu_{-1} \alpha^\nu_{-1} | 0_1; p \rangle + \cdots
\right]\,.
 \label{eq:field-expansion-2}
\end{equation}

Since the resulting action has an enormous gauge invariance given by
(\ref{eq:SFT-gauge}), it is often helpful to fix the gauge before
computing the action.  A particularly useful gauge choice is the
Feynman-Siegel gauge
\begin{equation}
b_0 | \Psi \rangle = 0\,.
\label{eq:FS-gauge}
\end{equation}
This is a good gauge choice
locally, fixing the linear gauge transformations $\delta | \Psi
\rangle = Q | \Lambda \rangle$.  This gauge choice is not, however,
globally valid; we will return to this point later.
In this gauge, all fields in the string field expansion which are
associated with states having an antighost zero-mode $c_0$ are taken
to vanish.  For example, the field $\chi (p)$ in
(\ref{eq:field-expansion-2}) vanishes.  In Feynman-Siegel gauge, the
BRST operator takes the simple form
\begin{equation}
Q = c_0L_0 =c_0 (N + p^2 -1)
\label{eq:FS-BRST}
\end{equation}
where $N$ is the total (matter + ghost) oscillator number.

Using (\ref{eq:FS-BRST}), it is straightforward to write the quadratic
terms in the string field action.  They are
\begin{equation}
\frac{1}{2}\langle V_2 | \Psi, Q \Psi \rangle =
\int d^{26} p \; \left\{
\phi (-p) \left[ \frac{p^2 -1}{2}  \right]\phi (p)
+ A_\mu (-p) \left[ \frac{p^2}{2}  \right]A^\mu (p) + \cdots
\right\}\,.
\end{equation}

The cubic part of the action can also be computed term by term,
although the terms are somewhat more complicated.  The leading terms
in the cubic action  are
given by
\begin{eqnarray}
\lefteqn{\frac{1}{3}
\langle V_3 | \Psi, \Psi, \Psi \rangle =}\label{eq:expanded-action} \\
&  & 
\int d^{26}pd^{26}q \; \frac{\kappa g}{3}  \;
 e^{(\ln 16/27) (p^2 + q^2 + p \cdot q)}
\left\{\phi (-p) \phi (-q) \phi (p + q) + \frac{16}{9}
A^{\mu} (-p) A_\mu (-q) \phi (p + q) 
 \right. 
\nonumber
\\
 & & \hspace{2.4in}
\left.
- \frac{8}{9}
(p^\mu +2q^\mu) (2p^{\nu} + q^{\nu})A^{\mu} (-p) A_\nu (-q) \phi (p +
q)  + \cdots
\right\} \nonumber
\end{eqnarray}
In computing the $\phi^3$ term we have used 
\begin{equation}
V^{rs}_{00} = \delta^{rs}
\ln (\frac{27}{16} )
\end{equation}
The $A^2 \phi$ term uses 
\begin{equation}
V^{rs}_{11} = -\frac{16}{27} , \; \;r \neq s,
\end{equation}
while the $ (A \cdot p)^2 \phi$ term uses 
\begin{equation}
V^{12}_{10} = -V^{13}_{10} = -\frac{2 \sqrt{2}}{3 \sqrt{3}} 
\end{equation}
The most striking feature of this action is that for a generic set
of three fields, there is a {\it nonlocal} cubic interaction term,
containing an exponential of a quadratic form in the momenta.  This
means that the target space formulation of string theory has a
dramatically different character from a standard quantum field theory.
From the point of view of quantum field theory, string field theory
seems to contain an infinite number of nonrenormalizable interactions.
Just like the simpler case of noncommutative field theories, however,
the magic of string theory seems to combine this infinite set of
interactions into a sensible model.  [Note, though, that we are
working here with the bosonic theory, which becomes problematic
quantum mechanically due to the closed string tachyon; the superstring
should be better behaved, although a complete understanding of
superstring field theory is still lacking despite recent progress
\cite{Aref'eva,Berkovits}].  For the purposes of the remainder of these
lectures, however, it will be sufficient for us to restrict attention
to the classical action at zero momentum, where the action
is quite well-behaved.


\subsection{General features of Witten's open bosonic SFT}
\label{sec:features}

There are several important aspects of Witten's open bosonic string
field theory which are worth reviewing here, although they will not be
central to the remainder of these lectures.

The first important aspect of this string field theory is that the
perturbative on-shell amplitudes computed using this SFT are in
precise agreement with the results of standard perturbative string
theory (CFT).  This result was shown by Giddings, Martinec, Witten,
and Zwiebach in \cite{Giddings-Martinec,gmw,Zwiebach-proof}; the basic idea
underlying this result is that in Feynman-Siegel gauge, the Feynman
diagrams of SFT precisely cover the appropriate moduli space of open
string diagrams of an arbitrary genus Riemann surface with boundaries,
with the ghost factors contributing the correct measure.  The
essential feature of this construction is the replacement of the
Feynman-Siegel gauge propagator $L_0^{-1}$ with a Schwinger
parameter
\begin{equation}
\frac{1}{L_0}  = \int_0^\infty dt \;e^{-tL_0}.
\label{eq:Schwinger}
\end{equation}
The Schwinger parameter $t$ plays the role of a modular parameter
measuring the length of the strip, for each propagator.  This sews the
string field theory diagram together into a Riemann surface for each
choice of Schwinger parameters; the result of
\cite{Giddings-Martinec,gmw,Zwiebach-proof} was to show that this
parameterization always precisely covers the moduli space correctly.
Thus, we know that to arbitrary orders in the string coupling the SFT
perturbative expansion agrees with standard string perturbation
theory, although  string field theory goes beyond the conformal
field theory approach since it is a nonperturbative, off-shell
formulation of the theory.

A consequence of the perturbative agreement between SFT and standard
perturbative string theory is that loop diagrams in open string field
theory must include closed string poles at appropriate values of the
external momenta.  It is well-known that while closed string theory in
a fixed space-time background (without D-branes) can be considered as
a complete and self-contained theory without including open strings,
the same is not true of open string theory.  Open strings can always
close up in virtual processes to form intermediate closed string
states.    The closed string poles were found explicitly in
the one-loop 2-point function of open string field theory in
 \cite{fgst}.  The appearance of these poles raises a very important
question for open string field theory, namely: Can closed strings
appear as asymptotic states in open string field theory?  Indeed,
standard arguments of unitarity would seem to imply that open string
field theory cannot be consistent at the quantum level unless open
strings can scatter into outgoing closed string states.  This question
becomes particularly significant in the context of Sen's tachyon
condensation conjectures, where we expect that all open string degrees
of freedom disappear from the theory in the nonperturbative locally
stable vacuum.  We will discuss this issue further in Section  8.


\section{Evidence for the Sen conjectures}
\label{sec:evidence}

Now that we have a more concrete understanding of how to carry out
calculations in open string field theory, we can address the
conjectures made by Sen regarding tachyon condensation.  In subsection
6.1, we discuss evidence for Sen's first conjecture, which states that
there exists a stable vacuum with energy density $-T_{25}$.  In
Subsection 6.2, we discuss physics in the stable vacuum and Sen's
third conjecture, which states that open strings decouple completely
from the theory in this vacuum.  There is also a large body of
evidence by now for Sen's second conjecture (see
\cite{Harvey-Kraus,djmt,Moeller-sz} for some of the early papers in
this direction), but due to time and space constraints we will not
cover this work here\footnote{A more  extensive summary of this work
will appear in \cite{Taylor-Zwiebach}}.


\subsection{Level truncation and the stable vacuum}
\label{sec:level-truncation-vacuum}

Sen's first conjecture states that the string field theory action
should lead to a nontrivial vacuum solution, with energy density
\begin{equation}
-T_{25} = -\frac{1}{2 \pi^2 g^2}  \,.
\end{equation}
In this subsection we discuss evidence for the validity of this conjecture.

\begin{figure}
\setlength{\unitlength}{0.240900pt}
\ifx\plotpoint\undefined\newsavebox{\plotpoint}\fi
\sbox{\plotpoint}{\rule[-0.200pt]{0.400pt}{0.400pt}}%
\begin{picture}(1500,900)(0,0)
\font\gnuplot=cmr10 at 10pt
\gnuplot
\sbox{\plotpoint}{\rule[-0.200pt]{0.400pt}{0.400pt}}%
\put(280.0,675.0){\rule[-0.200pt]{4.818pt}{0.400pt}}
\put(260,675){\makebox(0,0)[r]{0}}
\put(1419.0,675.0){\rule[-0.200pt]{4.818pt}{0.400pt}}
\put(280.0,332.0){\rule[-0.200pt]{4.818pt}{0.400pt}}
\put(260,332){\makebox(0,0)[r]{$V (\phi_0)$}}
\put(1419.0,332.0){\rule[-0.200pt]{4.818pt}{0.400pt}}
\put(280.0,175.0){\rule[-0.200pt]{4.818pt}{0.400pt}}
\put(260,175){\makebox(0,0)[r]{$-T_{25} $}}
\put(1419.0,175.0){\rule[-0.200pt]{4.818pt}{0.400pt}}
\put(598.0,82.0){\rule[-0.200pt]{0.400pt}{4.818pt}}
\put(598,41){\makebox(0,0){0}}
\put(598.0,840.0){\rule[-0.200pt]{0.400pt}{4.818pt}}
\put(1127.0,82.0){\rule[-0.200pt]{0.400pt}{4.818pt}}
\put(1127,41){\makebox(0,0){$\phi_0 = 1/3g\bar{\kappa}$}}
\put(1127.0,840.0){\rule[-0.200pt]{0.400pt}{4.818pt}}
\put(280.0,675.0){\rule[-0.200pt]{279.203pt}{0.400pt}}
\put(1401,638){\makebox(0,0)[l]{$\phi$}}
\put(480,841){\makebox(0,0)[l]{$V (\phi)$}}
\put(598.0,82.0){\rule[-0.200pt]{0.400pt}{187.420pt}}
\sbox{\plotpoint}{\rule[-0.400pt]{0.800pt}{0.800pt}}%
\put(1213,786){\makebox(0,0)[r]{Level (0, 0) approximation}}
\put(1233.0,786.0){\rule[-0.400pt]{24.090pt}{0.800pt}}
\put(280,156){\usebox{\plotpoint}}
\multiput(281.41,156.00)(0.511,1.892){17}{\rule{0.123pt}{3.067pt}}
\multiput(278.34,156.00)(12.000,36.635){2}{\rule{0.800pt}{1.533pt}}
\multiput(293.40,199.00)(0.512,1.935){15}{\rule{0.123pt}{3.109pt}}
\multiput(290.34,199.00)(11.000,33.547){2}{\rule{0.800pt}{1.555pt}}
\multiput(304.41,239.00)(0.511,1.666){17}{\rule{0.123pt}{2.733pt}}
\multiput(301.34,239.00)(12.000,32.327){2}{\rule{0.800pt}{1.367pt}}
\multiput(316.41,277.00)(0.511,1.621){17}{\rule{0.123pt}{2.667pt}}
\multiput(313.34,277.00)(12.000,31.465){2}{\rule{0.800pt}{1.333pt}}
\multiput(328.41,314.00)(0.511,1.485){17}{\rule{0.123pt}{2.467pt}}
\multiput(325.34,314.00)(12.000,28.880){2}{\rule{0.800pt}{1.233pt}}
\multiput(340.40,348.00)(0.512,1.536){15}{\rule{0.123pt}{2.527pt}}
\multiput(337.34,348.00)(11.000,26.755){2}{\rule{0.800pt}{1.264pt}}
\multiput(351.41,380.00)(0.511,1.304){17}{\rule{0.123pt}{2.200pt}}
\multiput(348.34,380.00)(12.000,25.434){2}{\rule{0.800pt}{1.100pt}}
\multiput(363.41,410.00)(0.511,1.259){17}{\rule{0.123pt}{2.133pt}}
\multiput(360.34,410.00)(12.000,24.572){2}{\rule{0.800pt}{1.067pt}}
\multiput(375.40,439.00)(0.512,1.237){15}{\rule{0.123pt}{2.091pt}}
\multiput(372.34,439.00)(11.000,21.660){2}{\rule{0.800pt}{1.045pt}}
\multiput(386.41,465.00)(0.511,1.078){17}{\rule{0.123pt}{1.867pt}}
\multiput(383.34,465.00)(12.000,21.126){2}{\rule{0.800pt}{0.933pt}}
\multiput(398.41,490.00)(0.511,0.988){17}{\rule{0.123pt}{1.733pt}}
\multiput(395.34,490.00)(12.000,19.402){2}{\rule{0.800pt}{0.867pt}}
\multiput(410.40,513.00)(0.512,0.988){15}{\rule{0.123pt}{1.727pt}}
\multiput(407.34,513.00)(11.000,17.415){2}{\rule{0.800pt}{0.864pt}}
\multiput(421.41,534.00)(0.511,0.807){17}{\rule{0.123pt}{1.467pt}}
\multiput(418.34,534.00)(12.000,15.956){2}{\rule{0.800pt}{0.733pt}}
\multiput(433.41,553.00)(0.511,0.762){17}{\rule{0.123pt}{1.400pt}}
\multiput(430.34,553.00)(12.000,15.094){2}{\rule{0.800pt}{0.700pt}}
\multiput(445.41,571.00)(0.511,0.717){17}{\rule{0.123pt}{1.333pt}}
\multiput(442.34,571.00)(12.000,14.233){2}{\rule{0.800pt}{0.667pt}}
\multiput(457.40,588.00)(0.512,0.639){15}{\rule{0.123pt}{1.218pt}}
\multiput(454.34,588.00)(11.000,11.472){2}{\rule{0.800pt}{0.609pt}}
\multiput(468.41,602.00)(0.511,0.536){17}{\rule{0.123pt}{1.067pt}}
\multiput(465.34,602.00)(12.000,10.786){2}{\rule{0.800pt}{0.533pt}}
\multiput(479.00,616.41)(0.491,0.511){17}{\rule{1.000pt}{0.123pt}}
\multiput(479.00,613.34)(9.924,12.000){2}{\rule{0.500pt}{0.800pt}}
\multiput(491.00,628.40)(0.489,0.512){15}{\rule{1.000pt}{0.123pt}}
\multiput(491.00,625.34)(8.924,11.000){2}{\rule{0.500pt}{0.800pt}}
\multiput(502.00,639.40)(0.674,0.516){11}{\rule{1.267pt}{0.124pt}}
\multiput(502.00,636.34)(9.371,9.000){2}{\rule{0.633pt}{0.800pt}}
\multiput(514.00,648.40)(0.913,0.526){7}{\rule{1.571pt}{0.127pt}}
\multiput(514.00,645.34)(8.738,7.000){2}{\rule{0.786pt}{0.800pt}}
\multiput(526.00,655.40)(0.913,0.526){7}{\rule{1.571pt}{0.127pt}}
\multiput(526.00,652.34)(8.738,7.000){2}{\rule{0.786pt}{0.800pt}}
\multiput(538.00,662.38)(1.432,0.560){3}{\rule{1.960pt}{0.135pt}}
\multiput(538.00,659.34)(6.932,5.000){2}{\rule{0.980pt}{0.800pt}}
\put(549,666.34){\rule{2.600pt}{0.800pt}}
\multiput(549.00,664.34)(6.604,4.000){2}{\rule{1.300pt}{0.800pt}}
\put(561,669.34){\rule{2.891pt}{0.800pt}}
\multiput(561.00,668.34)(6.000,2.000){2}{\rule{1.445pt}{0.800pt}}
\put(573,671.34){\rule{2.650pt}{0.800pt}}
\multiput(573.00,670.34)(5.500,2.000){2}{\rule{1.325pt}{0.800pt}}
\put(584,672.84){\rule{2.891pt}{0.800pt}}
\multiput(584.00,672.34)(6.000,1.000){2}{\rule{1.445pt}{0.800pt}}
\put(596,672.84){\rule{2.891pt}{0.800pt}}
\multiput(596.00,673.34)(6.000,-1.000){2}{\rule{1.445pt}{0.800pt}}
\put(608,671.84){\rule{2.891pt}{0.800pt}}
\multiput(608.00,672.34)(6.000,-1.000){2}{\rule{1.445pt}{0.800pt}}
\put(620,670.34){\rule{2.650pt}{0.800pt}}
\multiput(620.00,671.34)(5.500,-2.000){2}{\rule{1.325pt}{0.800pt}}
\put(631,667.84){\rule{2.891pt}{0.800pt}}
\multiput(631.00,669.34)(6.000,-3.000){2}{\rule{1.445pt}{0.800pt}}
\put(643,664.34){\rule{2.600pt}{0.800pt}}
\multiput(643.00,666.34)(6.604,-4.000){2}{\rule{1.300pt}{0.800pt}}
\multiput(655.00,662.06)(1.432,-0.560){3}{\rule{1.960pt}{0.135pt}}
\multiput(655.00,662.34)(6.932,-5.000){2}{\rule{0.980pt}{0.800pt}}
\multiput(666.00,657.07)(1.132,-0.536){5}{\rule{1.800pt}{0.129pt}}
\multiput(666.00,657.34)(8.264,-6.000){2}{\rule{0.900pt}{0.800pt}}
\multiput(678.00,651.07)(1.132,-0.536){5}{\rule{1.800pt}{0.129pt}}
\multiput(678.00,651.34)(8.264,-6.000){2}{\rule{0.900pt}{0.800pt}}
\multiput(690.00,645.08)(0.825,-0.526){7}{\rule{1.457pt}{0.127pt}}
\multiput(690.00,645.34)(7.976,-7.000){2}{\rule{0.729pt}{0.800pt}}
\multiput(701.00,638.08)(0.913,-0.526){7}{\rule{1.571pt}{0.127pt}}
\multiput(701.00,638.34)(8.738,-7.000){2}{\rule{0.786pt}{0.800pt}}
\multiput(713.00,631.08)(0.774,-0.520){9}{\rule{1.400pt}{0.125pt}}
\multiput(713.00,631.34)(9.094,-8.000){2}{\rule{0.700pt}{0.800pt}}
\multiput(725.00,623.08)(0.674,-0.516){11}{\rule{1.267pt}{0.124pt}}
\multiput(725.00,623.34)(9.371,-9.000){2}{\rule{0.633pt}{0.800pt}}
\multiput(737.00,614.08)(0.611,-0.516){11}{\rule{1.178pt}{0.124pt}}
\multiput(737.00,614.34)(8.555,-9.000){2}{\rule{0.589pt}{0.800pt}}
\multiput(748.00,605.08)(0.674,-0.516){11}{\rule{1.267pt}{0.124pt}}
\multiput(748.00,605.34)(9.371,-9.000){2}{\rule{0.633pt}{0.800pt}}
\multiput(760.00,596.08)(0.599,-0.514){13}{\rule{1.160pt}{0.124pt}}
\multiput(760.00,596.34)(9.592,-10.000){2}{\rule{0.580pt}{0.800pt}}
\multiput(772.00,586.08)(0.543,-0.514){13}{\rule{1.080pt}{0.124pt}}
\multiput(772.00,586.34)(8.758,-10.000){2}{\rule{0.540pt}{0.800pt}}
\multiput(783.00,576.08)(0.539,-0.512){15}{\rule{1.073pt}{0.123pt}}
\multiput(783.00,576.34)(9.774,-11.000){2}{\rule{0.536pt}{0.800pt}}
\multiput(795.00,565.08)(0.539,-0.512){15}{\rule{1.073pt}{0.123pt}}
\multiput(795.00,565.34)(9.774,-11.000){2}{\rule{0.536pt}{0.800pt}}
\multiput(807.00,554.08)(0.539,-0.512){15}{\rule{1.073pt}{0.123pt}}
\multiput(807.00,554.34)(9.774,-11.000){2}{\rule{0.536pt}{0.800pt}}
\multiput(819.00,543.08)(0.489,-0.512){15}{\rule{1.000pt}{0.123pt}}
\multiput(819.00,543.34)(8.924,-11.000){2}{\rule{0.500pt}{0.800pt}}
\multiput(830.00,532.08)(0.539,-0.512){15}{\rule{1.073pt}{0.123pt}}
\multiput(830.00,532.34)(9.774,-11.000){2}{\rule{0.536pt}{0.800pt}}
\multiput(842.00,521.08)(0.539,-0.512){15}{\rule{1.073pt}{0.123pt}}
\multiput(842.00,521.34)(9.774,-11.000){2}{\rule{0.536pt}{0.800pt}}
\multiput(855.40,507.55)(0.512,-0.539){15}{\rule{0.123pt}{1.073pt}}
\multiput(852.34,509.77)(11.000,-9.774){2}{\rule{0.800pt}{0.536pt}}
\multiput(865.00,498.08)(0.539,-0.512){15}{\rule{1.073pt}{0.123pt}}
\multiput(865.00,498.34)(9.774,-11.000){2}{\rule{0.536pt}{0.800pt}}
\multiput(877.00,487.08)(0.491,-0.511){17}{\rule{1.000pt}{0.123pt}}
\multiput(877.00,487.34)(9.924,-12.000){2}{\rule{0.500pt}{0.800pt}}
\multiput(889.00,475.08)(0.489,-0.512){15}{\rule{1.000pt}{0.123pt}}
\multiput(889.00,475.34)(8.924,-11.000){2}{\rule{0.500pt}{0.800pt}}
\multiput(900.00,464.08)(0.539,-0.512){15}{\rule{1.073pt}{0.123pt}}
\multiput(900.00,464.34)(9.774,-11.000){2}{\rule{0.536pt}{0.800pt}}
\multiput(912.00,453.08)(0.539,-0.512){15}{\rule{1.073pt}{0.123pt}}
\multiput(912.00,453.34)(9.774,-11.000){2}{\rule{0.536pt}{0.800pt}}
\multiput(924.00,442.08)(0.599,-0.514){13}{\rule{1.160pt}{0.124pt}}
\multiput(924.00,442.34)(9.592,-10.000){2}{\rule{0.580pt}{0.800pt}}
\multiput(936.00,432.08)(0.489,-0.512){15}{\rule{1.000pt}{0.123pt}}
\multiput(936.00,432.34)(8.924,-11.000){2}{\rule{0.500pt}{0.800pt}}
\multiput(947.00,421.08)(0.599,-0.514){13}{\rule{1.160pt}{0.124pt}}
\multiput(947.00,421.34)(9.592,-10.000){2}{\rule{0.580pt}{0.800pt}}
\multiput(959.00,411.08)(0.674,-0.516){11}{\rule{1.267pt}{0.124pt}}
\multiput(959.00,411.34)(9.371,-9.000){2}{\rule{0.633pt}{0.800pt}}
\multiput(971.00,402.08)(0.543,-0.514){13}{\rule{1.080pt}{0.124pt}}
\multiput(971.00,402.34)(8.758,-10.000){2}{\rule{0.540pt}{0.800pt}}
\multiput(982.00,392.08)(0.774,-0.520){9}{\rule{1.400pt}{0.125pt}}
\multiput(982.00,392.34)(9.094,-8.000){2}{\rule{0.700pt}{0.800pt}}
\multiput(994.00,384.08)(0.674,-0.516){11}{\rule{1.267pt}{0.124pt}}
\multiput(994.00,384.34)(9.371,-9.000){2}{\rule{0.633pt}{0.800pt}}
\multiput(1006.00,375.08)(0.913,-0.526){7}{\rule{1.571pt}{0.127pt}}
\multiput(1006.00,375.34)(8.738,-7.000){2}{\rule{0.786pt}{0.800pt}}
\multiput(1018.00,368.08)(0.700,-0.520){9}{\rule{1.300pt}{0.125pt}}
\multiput(1018.00,368.34)(8.302,-8.000){2}{\rule{0.650pt}{0.800pt}}
\multiput(1029.00,360.07)(1.132,-0.536){5}{\rule{1.800pt}{0.129pt}}
\multiput(1029.00,360.34)(8.264,-6.000){2}{\rule{0.900pt}{0.800pt}}
\multiput(1041.00,354.07)(1.132,-0.536){5}{\rule{1.800pt}{0.129pt}}
\multiput(1041.00,354.34)(8.264,-6.000){2}{\rule{0.900pt}{0.800pt}}
\multiput(1053.00,348.06)(1.432,-0.560){3}{\rule{1.960pt}{0.135pt}}
\multiput(1053.00,348.34)(6.932,-5.000){2}{\rule{0.980pt}{0.800pt}}
\put(1064,341.34){\rule{2.600pt}{0.800pt}}
\multiput(1064.00,343.34)(6.604,-4.000){2}{\rule{1.300pt}{0.800pt}}
\put(1076,337.34){\rule{2.600pt}{0.800pt}}
\multiput(1076.00,339.34)(6.604,-4.000){2}{\rule{1.300pt}{0.800pt}}
\put(1088,333.84){\rule{2.650pt}{0.800pt}}
\multiput(1088.00,335.34)(5.500,-3.000){2}{\rule{1.325pt}{0.800pt}}
\put(1099,331.84){\rule{2.891pt}{0.800pt}}
\multiput(1099.00,332.34)(6.000,-1.000){2}{\rule{1.445pt}{0.800pt}}
\put(1111,330.84){\rule{2.891pt}{0.800pt}}
\multiput(1111.00,331.34)(6.000,-1.000){2}{\rule{1.445pt}{0.800pt}}
\put(1135,330.84){\rule{2.650pt}{0.800pt}}
\multiput(1135.00,330.34)(5.500,1.000){2}{\rule{1.325pt}{0.800pt}}
\put(1146,332.34){\rule{2.891pt}{0.800pt}}
\multiput(1146.00,331.34)(6.000,2.000){2}{\rule{1.445pt}{0.800pt}}
\put(1158,335.34){\rule{2.600pt}{0.800pt}}
\multiput(1158.00,333.34)(6.604,4.000){2}{\rule{1.300pt}{0.800pt}}
\put(1170,339.34){\rule{2.400pt}{0.800pt}}
\multiput(1170.00,337.34)(6.019,4.000){2}{\rule{1.200pt}{0.800pt}}
\multiput(1181.00,344.39)(1.132,0.536){5}{\rule{1.800pt}{0.129pt}}
\multiput(1181.00,341.34)(8.264,6.000){2}{\rule{0.900pt}{0.800pt}}
\multiput(1193.00,350.40)(0.913,0.526){7}{\rule{1.571pt}{0.127pt}}
\multiput(1193.00,347.34)(8.738,7.000){2}{\rule{0.786pt}{0.800pt}}
\multiput(1205.00,357.40)(0.674,0.516){11}{\rule{1.267pt}{0.124pt}}
\multiput(1205.00,354.34)(9.371,9.000){2}{\rule{0.633pt}{0.800pt}}
\multiput(1217.00,366.40)(0.611,0.516){11}{\rule{1.178pt}{0.124pt}}
\multiput(1217.00,363.34)(8.555,9.000){2}{\rule{0.589pt}{0.800pt}}
\multiput(1228.00,375.41)(0.491,0.511){17}{\rule{1.000pt}{0.123pt}}
\multiput(1228.00,372.34)(9.924,12.000){2}{\rule{0.500pt}{0.800pt}}
\multiput(1240.00,387.41)(0.491,0.511){17}{\rule{1.000pt}{0.123pt}}
\multiput(1240.00,384.34)(9.924,12.000){2}{\rule{0.500pt}{0.800pt}}
\multiput(1253.40,398.00)(0.512,0.639){15}{\rule{0.123pt}{1.218pt}}
\multiput(1250.34,398.00)(11.000,11.472){2}{\rule{0.800pt}{0.609pt}}
\multiput(1264.41,412.00)(0.511,0.671){17}{\rule{0.123pt}{1.267pt}}
\multiput(1261.34,412.00)(12.000,13.371){2}{\rule{0.800pt}{0.633pt}}
\multiput(1276.41,428.00)(0.511,0.717){17}{\rule{0.123pt}{1.333pt}}
\multiput(1273.34,428.00)(12.000,14.233){2}{\rule{0.800pt}{0.667pt}}
\multiput(1288.41,445.00)(0.511,0.807){17}{\rule{0.123pt}{1.467pt}}
\multiput(1285.34,445.00)(12.000,15.956){2}{\rule{0.800pt}{0.733pt}}
\multiput(1300.40,464.00)(0.512,0.938){15}{\rule{0.123pt}{1.655pt}}
\multiput(1297.34,464.00)(11.000,16.566){2}{\rule{0.800pt}{0.827pt}}
\multiput(1311.41,484.00)(0.511,0.943){17}{\rule{0.123pt}{1.667pt}}
\multiput(1308.34,484.00)(12.000,18.541){2}{\rule{0.800pt}{0.833pt}}
\multiput(1323.41,506.00)(0.511,1.033){17}{\rule{0.123pt}{1.800pt}}
\multiput(1320.34,506.00)(12.000,20.264){2}{\rule{0.800pt}{0.900pt}}
\multiput(1335.40,530.00)(0.512,1.237){15}{\rule{0.123pt}{2.091pt}}
\multiput(1332.34,530.00)(11.000,21.660){2}{\rule{0.800pt}{1.045pt}}
\multiput(1346.41,556.00)(0.511,1.169){17}{\rule{0.123pt}{2.000pt}}
\multiput(1343.34,556.00)(12.000,22.849){2}{\rule{0.800pt}{1.000pt}}
\multiput(1358.41,583.00)(0.511,1.304){17}{\rule{0.123pt}{2.200pt}}
\multiput(1355.34,583.00)(12.000,25.434){2}{\rule{0.800pt}{1.100pt}}
\multiput(1370.40,613.00)(0.512,1.486){15}{\rule{0.123pt}{2.455pt}}
\multiput(1367.34,613.00)(11.000,25.905){2}{\rule{0.800pt}{1.227pt}}
\multiput(1381.41,644.00)(0.511,1.440){17}{\rule{0.123pt}{2.400pt}}
\multiput(1378.34,644.00)(12.000,28.019){2}{\rule{0.800pt}{1.200pt}}
\multiput(1393.41,677.00)(0.511,1.530){17}{\rule{0.123pt}{2.533pt}}
\multiput(1390.34,677.00)(12.000,29.742){2}{\rule{0.800pt}{1.267pt}}
\multiput(1405.41,712.00)(0.511,1.666){17}{\rule{0.123pt}{2.733pt}}
\multiput(1402.34,712.00)(12.000,32.327){2}{\rule{0.800pt}{1.367pt}}
\multiput(1417.40,750.00)(0.512,1.885){15}{\rule{0.123pt}{3.036pt}}
\multiput(1414.34,750.00)(11.000,32.698){2}{\rule{0.800pt}{1.518pt}}
\multiput(1428.41,789.00)(0.511,1.847){17}{\rule{0.123pt}{3.000pt}}
\multiput(1425.34,789.00)(12.000,35.773){2}{\rule{0.800pt}{1.500pt}}
\put(1123.0,332.0){\rule[-0.400pt]{2.891pt}{0.800pt}}
\sbox{\plotpoint}{\rule[-0.500pt]{1.000pt}{1.000pt}}%
\put(1213,745){\makebox(0,0)[r]{Level (2, 6) approximation}}
\multiput(1233,745)(20.756,0.000){5}{\usebox{\plotpoint}}
\put(1333,745){\usebox{\plotpoint}}
\put(399,477){\usebox{\plotpoint}}
\multiput(399,477)(7.708,19.271){2}{\usebox{\plotpoint}}
\put(415.53,515.06){\usebox{\plotpoint}}
\put(424.91,533.57){\usebox{\plotpoint}}
\put(434.87,551.78){\usebox{\plotpoint}}
\put(445.52,569.59){\usebox{\plotpoint}}
\put(457.19,586.74){\usebox{\plotpoint}}
\put(470.16,602.94){\usebox{\plotpoint}}
\put(483.09,619.09){\usebox{\plotpoint}}
\put(498.01,633.51){\usebox{\plotpoint}}
\put(514.36,646.27){\usebox{\plotpoint}}
\put(531.74,657.59){\usebox{\plotpoint}}
\put(550.57,666.21){\usebox{\plotpoint}}
\put(570.43,672.11){\usebox{\plotpoint}}
\put(590.95,675.00){\usebox{\plotpoint}}
\put(611.66,674.17){\usebox{\plotpoint}}
\put(632.00,670.25){\usebox{\plotpoint}}
\put(651.89,664.42){\usebox{\plotpoint}}
\put(671.22,656.89){\usebox{\plotpoint}}
\put(689.54,647.17){\usebox{\plotpoint}}
\put(707.14,636.16){\usebox{\plotpoint}}
\put(724.30,624.53){\usebox{\plotpoint}}
\put(740.53,611.60){\usebox{\plotpoint}}
\put(756.59,598.48){\usebox{\plotpoint}}
\put(772.21,584.82){\usebox{\plotpoint}}
\put(787.36,570.64){\usebox{\plotpoint}}
\put(802.52,556.48){\usebox{\plotpoint}}
\put(817.19,541.81){\usebox{\plotpoint}}
\put(831.36,526.64){\usebox{\plotpoint}}
\put(845.91,511.85){\usebox{\plotpoint}}
\put(859.70,496.34){\usebox{\plotpoint}}
\put(873.43,480.77){\usebox{\plotpoint}}
\put(887.21,465.26){\usebox{\plotpoint}}
\put(901.00,449.75){\usebox{\plotpoint}}
\put(914.79,434.23){\usebox{\plotpoint}}
\put(928.58,418.72){\usebox{\plotpoint}}
\put(942.37,403.21){\usebox{\plotpoint}}
\put(956.16,387.69){\usebox{\plotpoint}}
\put(969.95,372.18){\usebox{\plotpoint}}
\put(984.22,357.12){\usebox{\plotpoint}}
\put(997.66,341.34){\usebox{\plotpoint}}
\put(1012.19,326.53){\usebox{\plotpoint}}
\put(1026.53,311.54){\usebox{\plotpoint}}
\put(1041.66,297.34){\usebox{\plotpoint}}
\put(1056.74,283.10){\usebox{\plotpoint}}
\put(1072.36,269.44){\usebox{\plotpoint}}
\put(1088.36,256.23){\usebox{\plotpoint}}
\put(1104.96,243.78){\usebox{\plotpoint}}
\put(1121.72,231.55){\usebox{\plotpoint}}
\put(1139.45,220.77){\usebox{\plotpoint}}
\put(1158.02,211.49){\usebox{\plotpoint}}
\put(1176.94,203.02){\usebox{\plotpoint}}
\put(1197.07,197.98){\usebox{\plotpoint}}
\put(1217.57,194.93){\usebox{\plotpoint}}
\put(1238.23,194.65){\usebox{\plotpoint}}
\put(1258.75,197.69){\usebox{\plotpoint}}
\put(1278.64,203.49){\usebox{\plotpoint}}
\put(1297.20,212.75){\usebox{\plotpoint}}
\put(1314.64,223.98){\usebox{\plotpoint}}
\put(1330.58,237.26){\usebox{\plotpoint}}
\put(1345.49,251.68){\usebox{\plotpoint}}
\put(1359.28,267.19){\usebox{\plotpoint}}
\put(1371.94,283.62){\usebox{\plotpoint}}
\put(1383.55,300.82){\usebox{\plotpoint}}
\multiput(1391,312)(10.878,17.677){2}{\usebox{\plotpoint}}
\put(1415.61,354.15){\usebox{\plotpoint}}
\put(1425.26,372.53){\usebox{\plotpoint}}
\put(1434.38,391.17){\usebox{\plotpoint}}
\put(1439,401){\usebox{\plotpoint}}
\sbox{\plotpoint}{\rule[-0.200pt]{0.400pt}{0.400pt}}%
\put(1213,704){\makebox(0,0)[r]{Exact value of $-T_{25}$}}
\multiput(1233,704)(20.756,0.000){5}{\usebox{\plotpoint}}
\put(1333,704){\usebox{\plotpoint}}
\put(280,175){\usebox{\plotpoint}}
\put(280.00,175.00){\usebox{\plotpoint}}
\put(300.76,175.00){\usebox{\plotpoint}}
\put(321.51,175.00){\usebox{\plotpoint}}
\put(342.27,175.00){\usebox{\plotpoint}}
\put(363.02,175.00){\usebox{\plotpoint}}
\put(383.78,175.00){\usebox{\plotpoint}}
\put(404.53,175.00){\usebox{\plotpoint}}
\put(425.29,175.00){\usebox{\plotpoint}}
\put(446.04,175.00){\usebox{\plotpoint}}
\put(466.80,175.00){\usebox{\plotpoint}}
\put(487.55,175.00){\usebox{\plotpoint}}
\put(508.31,175.00){\usebox{\plotpoint}}
\put(529.07,175.00){\usebox{\plotpoint}}
\put(549.82,175.00){\usebox{\plotpoint}}
\put(570.58,175.00){\usebox{\plotpoint}}
\put(591.33,175.00){\usebox{\plotpoint}}
\put(612.09,175.00){\usebox{\plotpoint}}
\put(632.84,175.00){\usebox{\plotpoint}}
\put(653.60,175.00){\usebox{\plotpoint}}
\put(674.35,175.00){\usebox{\plotpoint}}
\put(695.11,175.00){\usebox{\plotpoint}}
\put(715.87,175.00){\usebox{\plotpoint}}
\put(736.62,175.00){\usebox{\plotpoint}}
\put(757.38,175.00){\usebox{\plotpoint}}
\put(778.13,175.00){\usebox{\plotpoint}}
\put(798.89,175.00){\usebox{\plotpoint}}
\put(819.64,175.00){\usebox{\plotpoint}}
\put(840.40,175.00){\usebox{\plotpoint}}
\put(861.15,175.00){\usebox{\plotpoint}}
\put(881.91,175.00){\usebox{\plotpoint}}
\put(902.66,175.00){\usebox{\plotpoint}}
\put(923.42,175.00){\usebox{\plotpoint}}
\put(944.18,175.00){\usebox{\plotpoint}}
\put(964.93,175.00){\usebox{\plotpoint}}
\put(985.69,175.00){\usebox{\plotpoint}}
\put(1006.44,175.00){\usebox{\plotpoint}}
\put(1027.20,175.00){\usebox{\plotpoint}}
\put(1047.95,175.00){\usebox{\plotpoint}}
\put(1068.71,175.00){\usebox{\plotpoint}}
\put(1089.46,175.00){\usebox{\plotpoint}}
\put(1110.22,175.00){\usebox{\plotpoint}}
\put(1130.98,175.00){\usebox{\plotpoint}}
\put(1151.73,175.00){\usebox{\plotpoint}}
\put(1172.49,175.00){\usebox{\plotpoint}}
\put(1193.24,175.00){\usebox{\plotpoint}}
\put(1214.00,175.00){\usebox{\plotpoint}}
\put(1234.75,175.00){\usebox{\plotpoint}}
\put(1255.51,175.00){\usebox{\plotpoint}}
\put(1276.26,175.00){\usebox{\plotpoint}}
\put(1297.02,175.00){\usebox{\plotpoint}}
\put(1317.77,175.00){\usebox{\plotpoint}}
\put(1338.53,175.00){\usebox{\plotpoint}}
\put(1359.29,175.00){\usebox{\plotpoint}}
\put(1380.04,175.00){\usebox{\plotpoint}}
\put(1400.80,175.00){\usebox{\plotpoint}}
\put(1421.55,175.00){\usebox{\plotpoint}}
\put(1439,175){\usebox{\plotpoint}}
\sbox{\plotpoint}{\rule[-0.500pt]{1.000pt}{1.000pt}}%
\put(1127,332){\raisebox{-.8pt}{\circle{15}}}
\put(1229,194){\raisebox{-.8pt}{\circle{15}}}
\put(396,471){\raisebox{-.8pt}{\circle*{15}}}
\end{picture}
\caption[x]{\footnotesize The effective tachyon potential in level (0,
  0) and (2, 6) truncations.  The open circles denote minima in each
  level truncation.  The filled circle denotes a branch point where
  the level (2, 6) truncation approximation reaches the limit of
  Feynman-Siegel gauge validity.}
\label{f:potential}
\end{figure}
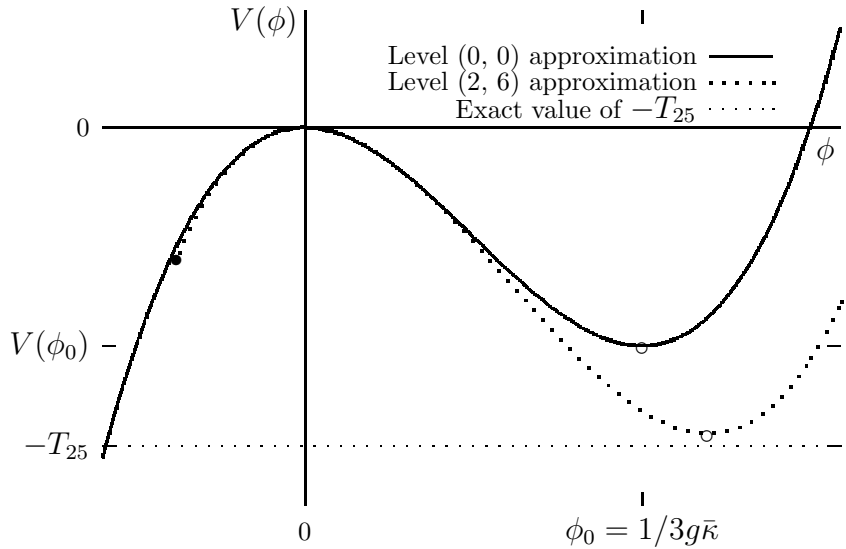

The string field theory equation of motion is
\begin{equation}
Q \Psi + g \Psi \star \Psi = 0 \,.
\label{eq:SFT-EOM}
\end{equation}
Despite much work over the last few years, there
is still no analytic solution of this equation of motion\footnote{as
of January, 2003}.  There is, however, a systematic approximation
scheme, known as level truncation, which can be used to solve this
equation numerically.  The level $(L, I) $ truncation of the full
string field theory involves dropping all fields at level $N > L$, and
disregarding any cubic interaction terms between fields whose total level is
greater than $I$.  For example, the simplest truncation of the theory
is the level (0, 0) truncation.  Including only $p = 0$ components of
the tachyon field, with the justification that we are looking for a
Lorentz-invariant vacuum, the theory in this truncation is simply
described by a potential for the tachyon zero-mode
\begin{equation}
V (\phi) = -\frac{1}{2}\phi^2 + g \bar{\kappa} \phi^3 \,.
\end{equation}
where $\bar{\kappa} = \kappa/3 = 3^{7/2}/2^6$.
This cubic function is graphed in Figure~\ref{f:potential}.
Clearly, this potential has a local minimum at
\begin{equation}
\phi_0 = \frac{1}{3g \bar{\kappa}}  \,.
\end{equation}
At this point the potential is
\begin{equation}
V (\phi_0) = -\frac{1}{54}  \frac{1}{g^2 \bar{\kappa}^2}  =
-\frac{2^{11}}{3^{10}}  \frac{1}{g^2} 
\approx (0.68) \left( -\frac{1}{2 \pi^2 g^2}  \right)
\end{equation}
Thus, we see that simply including the tachyon zero-mode gives a
nontrivial vacuum with $68\%$ of the vacuum energy density predicted
by Sen.  This vacuum is denoted by an open circle in Figure~\ref{f:potential}.

At higher levels of truncation, there are a multitude of fields with
various tensor structures.  However, again assuming that we are
looking for a vacuum which preserves Lorentz symmetry, we can restrict
attention to the interactions between scalar fields at $p = 0$.  We
will work in Feynman-Siegel gauge to simplify calculations.  The
situation is further simplified by the existence of the ``twist''
symmetry mentioned in Section 5.3, which guarantees that no cubic
vertex between $p = 0$ scalar fields can connect three fields with a
total level which is odd.  This means that odd fields are not relevant
to diagrams with only external tachyons at tree level.  Thus, we need
only consider even-level scalar fields in looking for
Lorentz-preserving solutions to the SFT equations of motion.  With
these simplifications, in a general level truncation the string field
is simply expressed as a sum of a finite number of terms
\begin{equation}
\Psi_{\rm s} = \sum_{i} \phi_i | s_i \rangle
\label{eq:scalar-expansion}
\end{equation}
where $\phi_i$ are the zero-modes of the scalar fields associated with
even-level states $|s_i \rangle$.  For example, including fields up to
level 2, we have
\begin{equation}
\Psi_{{\rm s}} = \phi| 0_1 \rangle+
B \; (\alpha_{-1} \cdot \alpha_{-1})| 0_1 \rangle+
\beta \; b_{-1} c_{-1}| 0_1 \rangle\,.
\end{equation}
The potential for all the scalars appearing in the level-truncated
expansion (\ref{eq:scalar-expansion}) can be simply expressed as a
cubic polynomial in the zero-modes of the scalar fields
\begin{equation}
V = \sum_{i, j}d_{ij} \phi_i \phi_j + g \bar{\kappa}
\sum_{i, j, k}t_{ijk} \phi_i \phi_j \phi_k \,.
\label{eq:scalar-potential}
\end{equation}
Using the expressions for the Neumann coefficients given in Section
5.3, the potential for all the scalar fields up to level $L$ can be
computed in a level  $(L, I)$ truncation.  For example, the potential
in the level $(2, 6)$ truncation is given by
\begin{eqnarray}
V & = &  -\frac{1}{2}\phi^2 + 26  B^2 -\frac{1}{2}\beta^2
\nonumber\\
 &  &+ \bar{\kappa} g\left[ 
\phi^3  -\frac{130}{9} \phi^2 B -\frac{11}{9}  \phi^2 \beta
+ \frac{30212}{243}  \phi B^2
+ \frac{2860}{243}  \phi B \beta
+ \frac{19}{81}  \phi \beta^2
 \right. \label{eq:v26} \\
& &\hspace*{0.4in}\left.
-\frac{2178904}{6561}  B^3
-\frac{332332}{6561}  B^2 \beta
-\frac{2470}{2187}  B \beta^2
-\frac{1}{81}  \beta^3
 \right] \nonumber
\end{eqnarray}
As an example of how these terms arise, consider the $\phi^2 B$ term.
The coefficient in this term is given by
\begin{eqnarray}
g\, \langle V_3 | (| 0_1 \rangle \otimes | 0_1 \rangle \otimes
\alpha_{-1} \cdot \alpha_{-1} | 0_1 \rangle) & = & 
-g \bar{\kappa} \;  (3 \cdot 26) \; V^{11}_{11}\\
 & = & -g \bar{\kappa} \frac{130}{9}  \nonumber
\end{eqnarray}
where we have used $V^{11}_{11} = 5/27$.

In the level (2, 6) truncation of the theory, with potential
(\ref{eq:v26}), the nontrivial vacuum is found by
simultaneously solving the three quadratic equations found by setting
to zero the derivatives of (\ref{eq:v26}) with respect to
$\phi, B,$ and $\beta$.  There are a number of different solutions to
these equations, but only one is in the vicinity of $\phi = 1/3g
\bar{\kappa}$.  The solution of interest is
\begin{eqnarray}
\phi & \approx & 0.39766\, \frac{1}{g \bar{\kappa}}  \nonumber\\
B  &  \approx& 0.02045\, \frac{1}{g \bar{\kappa}}\\
\beta  &  \approx & -0.13897\, \frac{1}{g \bar{\kappa}} \nonumber
\end{eqnarray}
Plugging these values into the potential gives
\begin{equation}
E_{(2, 6)} = -0.95938 \,T_{25} \,,
\end{equation}
or 95.9\% of the result predicted by Sen.
This vacuum is denoted by an open circle in Figure~\ref{f:potential}.

It is a straightforward, although computationally intensive, project
to generalize this calculation to higher levels of truncation.  This
calculation was carried out to level (4, 8) by Kostelecky and Samuel
\cite{ks-open} many years ago.  They noted that the vacuum seemed to
be converging, but they lacked any physical picture giving meaning to
this vacuum.  Following Sen's conjectures, the level (4, 8)
calculation was done again using somewhat different methods by Sen and
Zwiebach \cite{Sen-Zwiebach}, who showed that the energy at this level
is $-0.986 \;T_{25}$.  The calculation was automated by Moeller and
Taylor \cite{Moeller-Taylor}, who calculated up to level (10, 20),
where there are 252 scalar fields.  Up to this level, the vacuum
energy converges monotonically, as shown in Table 1.
\begin{table}[htp]
\begin{center}
\begin{tabular}{|| c  || c |  c ||}
\hline
\hline
level & $ g \bar{\kappa}\langle \phi \rangle$ & $V/T_{25}$\\
\hline
\hline
(0, 0) & 0.3333 & -0.68462\\
\hline
(2, 4) & 0.3957 &  -0.94855\\
(2, 6) & 0.3977 &  -0.95938\\
\hline
(4, 8) & 0.4005 &  -0.98640\\
(4, 12) & 0.4007 & -0.98782\\
\hline
(6, 12) & 0.4004 & -0.99514\\
(6, 18) & 0.4004 & -0.99518\\
\hline
(8, 16) & 0.3999 & -0.99777\\
(8, 20) & 0.3997 & -0.99793\\
\hline
(10, 20) & 0.3992 & -0.99912\\
\hline
\hline
\end{tabular}
\caption[x]{\footnotesize Tachyon VEV and vacuum energy in stable
vacua of level-truncated theory}
\label{t:vacuum}
\end{center}
\end{table}
These numerical calculations indicate that level truncation of string
field theory leads to a good systematic approximation scheme for
computing the nonperturbative tachyon vacuum
\footnote{These were the best values for the vacuum
energy and effective potential
at the time of the lectures.  At strings 2002, Gaiotto and
Rastelli reported results up to level (18, 54)
 \cite{Gaiotto-Rastelli-strings}.  They found the surprising result
that while the energy monotonically approaches $-T_{25}$ up to level
12, at level (14, 42) the energy becomes $-1.0002\, T_{25}$, and that
the energy continues to decrease, reaching $-1.0005\, T_{25}$ at level
(18, 54).  In \cite{WT-Pade}, it was shown that this calculation could
be theoretically
extrapolated to higher levels using the result found in
 \cite{WT-perturbative} that perturbative amplitudes converge in level
truncation with errors described by a power series in $1/L$.  This
extrapolation suggests that the energy turns around again near $L
=28$, and then increases again, asymptotically approaching $-T_{25}$
as $L \rightarrow \infty$.  Further analysis supporting this
conclusion was given in \cite{gr-analysis}, where the effective
tachyon potential was extrapolated to higher order using results
calculated up to level 18.}.

It is interesting to consider the tachyon condensation problem from
the point of view of the effective tachyon potential.  If instead of
trying to solve the quadratic equations for all $N$ of the fields
appearing in (\ref{eq:scalar-potential}), we instead fix the tachyon
field $\phi$ and solve the quadratic equations for the remaining $N
-1$ fields, we can determine a effective potential $V (\phi)$ for the
tachyon field.  This was done numerically up to level $(10, 20)$ in
\cite{Moeller-Taylor}${}^6$.  At each level, the tachyon effective potential
smoothly interpolates between the perturbative vacuum and the
nonperturbative vacuum near $\phi = 0.4/g \bar{\kappa}$.  For example,
the tachyon effective potential at level (2, 6) is graphed in
Figure~\ref{f:potential}.  In all level truncations other than (0, 0)
and (2, 4), the tachyon effective potential has two branch point
singularities at which the continuous solution for the other fields
breaks down; for the level (2, 6) truncation, these branch points
occur at $\phi \approx -0.127/g \bar{\kappa}$ and $\phi \approx
2.293/g \bar{\kappa}$; the lower branch point is denoted by a solid
circle in Figure~\ref{f:potential}.  As a result of these branch
points, the tachyon effective potential is only valid for a finite
range of $\phi$, ranging between approximately $-0.1/g \bar{\kappa}$
and $ 0.6/g \bar{\kappa}$.  In \cite{Ellwood-Taylor-gauge} it was
demonstrated numerically that the branch points in the tachyon
effective potential arise because the trajectory in field space
associated with this potential encounters the boundary of the region
of Feynman-Siegel gauge validity.  As mentioned earlier,
Feynman-Siegel gauge is only valid in a finite-size region around the
perturbative vacuum.  It seems almost to be a fortunate accident that
the nonperturbative vacuum lies within the region of validity of this
gauge choice.  It is also worth mentioning here that in the
``background-independent'' formulation of SFT, the tachyon potential
can be computed exactly \cite{BSFT}.  In this formulation, there is no
branch point in the effective potential, which is unbounded below for
negative values of the tachyon.  On the other hand, the nontrivial
vacuum in the background-independent approach arises only as the
tachyon field goes to infinity, so it is harder to study the physics
of the stable vacuum from this point of view.

Another interesting perspective on the tachyon effective potential is
found by performing a perturbative computation of the coefficients in
this effective potential in the level-truncated theory.  
This gives a power series expansion of the effective tachyon potential
\begin{eqnarray}
V (\phi)  & = & \sum_{n = 2}^{ \infty}  c_n (\bar{\kappa} g)^{n-2} \phi^n
\label{eq:v}\\
& = & -\frac{1}{2}\phi^2 + (\bar{\kappa} g) \phi^3 + c_4 (\bar{\kappa} g)^2
\phi^4 + c_5 (\bar{\kappa} g)^3 \phi^5 +\cdots\nonumber
\end{eqnarray}
In \cite{Moeller-Taylor}, the coefficients up to $c_{60}$ were
computed in the level truncations up to (10, 20).  Because of the
branch point singularity near $\phi = -0.1/g \bar{\kappa}$, this
series has a radius of convergence much smaller than the value of
$\phi$ at the nonperturbative vacuum.  Thus, the energy at the stable
vacuum lies outside the naive range of perturbation theory\footnote{In
 \cite{WT-Pade}, however, it was shown that the method of Pad\'e
approximants enables us to compute the vacuum energy to excellent
precision given a reasonably small number of the coefficients $c_n$.
Thus, the stable vacuum is in some sense accessible from purely
perturbative calculations.}.


\subsection{Physics in the stable vacuum}
\label{sec:vacuum-physics}

We have seen that numerical results from level-truncated string field
theory strongly suggest the existence of a classically stable vacuum solution
$\Psi_0$ to the string field theory equation of motion
(\ref{eq:SFT-EOM}).  The state $\Psi_0$, while still unknown
analytically, has been determined numerically to a high degree of
precision.  This state seems like a very well-behaved string field
configuration.  While there is no positive-definite inner product on
the string field Fock space, the state $\Psi_0$ certainly has finite
norm under the natural inner product $\langle V_2| \Psi_0, c_0L_0
\Psi_0\rangle$, and is even better behaved under the product
$\langle V_2| \Psi_0, c_0
\Psi_0\rangle$.  Thus, it is natural to assume that $\Psi_0$ defines a
classically stable vacuum for the theory, around which we can expand
the action to find a new ``vacuum string field theory''.  Expanding
\begin{equation}
\Psi = \Psi_0 + \tilde{\Psi},
\end{equation}
we get the action
\begin{equation}
\tilde{S} (\tilde{\Psi}) = S (\Psi_0 + \tilde{ \Psi})
= S_0 -\frac{1}{2}\int \tilde{\Psi} \star  \tilde{Q} \tilde{\Psi} -\frac{g}{3}
\int \tilde{\Psi} \star 
\tilde{\Psi} \star \tilde{\Psi}\,.
\label{eq:VSFT-action}
\end{equation}
where
\begin{equation}
\tilde{Q} \Phi = Q \Phi + g (\Psi_0 \star \Phi + \Phi \star \Psi_0) \,.
\label{eq:VSFT-q}
\end{equation}
This string field theory around the stable vacuum has precisely the
same form as Witten's original cubic string field theory, only with a
different BRST operator $\tilde{Q}$, which so far is only determined
numerically.  Note that this formulation of vacuum string field theory
is distinct from the VSFT model of Rastelli, Sen, and Zwiebach (RSZ)
 \cite{RSZ}.  These authors make an Ansatz that the BRST operator takes a
pure ghost form, along the lines of $Q \rightarrow c_0$, and they
conjecture that the theory with such a BRST operator is equivalent to
the VSFT model given by the BRST operator (\ref{eq:VSFT-q}).  We
discuss the RSZ model again briefly in the next section.

Sen's third conjecture states that there should be no open string
excitations of the theory around $\Psi = \Psi_0$.  This implies that
there should be no solutions of the linearized equation $\tilde{Q}
\tilde{ \Psi}$ in the VSFT (\ref{eq:VSFT-action}) other than pure gauge
states of the form $\tilde{\Psi} = \tilde{Q} \tilde{\Lambda}$.  In
this subsection we discuss evidence for this conjecture.  

It may seem surprising to imagine that {\it all} the perturbative open
string degrees of freedom will vanish at a particular point in field
space, since these are all the degrees of freedom available in the
theory.  This is not a familiar phenomenon from quantum field theory.  To
understand how the open strings can decouple, it may be helpful to
begin by considering the simple example of the (0, 0) level-truncated
theory.  In this theory, the quadratic terms in the action become
\begin{equation}
-\int d^{26} p \;
\phi (-p) \left[ \frac{p^2 -1}{2}  + g \bar{\kappa} \left(
  \frac{16}{27}  \right)^{p^2} \cdot 3 \langle \phi \rangle \right]
\phi (p) \,.
\end{equation}
Taking
$\langle \phi \rangle = 1/3 \bar{\kappa} g$, we find that the
quadratic term is a transcendental expression which does not vanish
for any real value of $p^2$.  Thus, this theory has no poles, and the
tachyon has decoupled from the theory.  Of course, this is not the
full story, as there are still finite complex poles.  It does, however
suggest a mechanism by which the nonlocal parts of the action (encoded
in the exponential of $p^2$) can remove physical poles.

To get the full story, it is necessary to continue the analysis to
higher level.  At level 2, there are 7 scalar fields, 
the tachyon and the 6 fields associated with the Fock space states
\begin{eqnarray}
(\alpha_{-1} \cdot \alpha_{-1})| 0_1, p \rangle &  &
b_{-1} \cdot c_{-1}| 0_1, p \rangle \nonumber\\
c_{0} \cdot  b_{-1}| 0_1, p \rangle &  &
(p \cdot \alpha_{-2})| 0_1, p \rangle \\
(p \cdot \alpha_{-1})^2| 0_1, p \rangle &  &
(p \cdot  \alpha_{-1}) c_0b_1| 0_1, p \rangle  \nonumber
\end{eqnarray}
Note that in this analysis we cannot fix Feynman-Siegel gauge, as we
only believe that this gauge is valid for the zero-modes of the scalar
fields in the vacuum $\Psi_0$.  An attempt at analyzing the spectrum
of the theory in 
Feynman-Siegel gauge using level truncation was made in
 \cite{ks-open}, with no sensible results.  Diagonalizing the quadratic
term in the action on the full set of 7 fields of level $\leq 2$, we find that
poles develop at $M^2 = 0.9$ and $M^2 = 2.0$ (in string units, where
the tachyon has $M^2 = -1$) \cite{Ellwood-Taylor-spectrum}.  These
poles correspond to states satisfying $\tilde{Q} \tilde{\Psi} = 0$.
The question now is, are these states physical?  If they are exact
states, of the form $\tilde{\Psi} = \tilde{Q} \tilde{\Lambda}$, then
they are simply gauge degrees of freedom.  If not, however, then they
are states in the cohomology of $\tilde{Q}$ and should be associated
with physical degrees of freedom.  Unfortunately, we cannot precisely
determine whether the poles we find in level truncation are due to
exact states, as the level-truncation procedure breaks the condition
$\tilde{Q}^2 = 0$.  Thus, we can only measure {\it approximately}
whether a state is exact.  A detailed analysis of this question was
carried out in \cite{Ellwood-Taylor-spectrum}.  In this paper, all
terms in the SFT action of the form $\phi_i \; \psi_j (p) \; \psi_k (-p)$
were determined, where $\phi_i$ is a scalar zero-mode, and $\psi_{j,
k}$ are nonzero-momentum scalars.  In addition, all gauge
transformations involving at least one zero-momentum field were
computed up to level (6, 12).  At each level up to $L = 6$, the ghost
number 1 states in the kernel ${\rm Ker} \;\tilde{Q}^{(1)}_{(L, 2L)}$
were computed.  The extent to which each of these states lies in the
exact subspace was measured using the formula
\begin{equation}
\% \;{\rm exactness} = \sum_{i}\frac{(s \cdot e_i)^2}{ (s \cdot s)} 
\end{equation}
where $\{e_i\}$ are an orthonormal basis for ${\rm Im} \;
\tilde{Q}^{(0)}_{ (L, 2L)}$, the image of $\tilde{Q}$ acting on the
space of ghost number 0 states in the appropriate level truncation.
(Note that this measure involves a choice of inner product on the Fock
space; several natural inner products were tried, giving roughly
equivalent results).  The result of this analysis was that up to the
mass scale of the level truncation, $M^2 \leq L -1$, all the states in
the kernel of $\tilde{Q}^{(1)}$ were $\geq 99.9\%$ within the exact
subspace, for $L \geq 4$.  This result seems to give very strong
evidence for Sen's third conjecture that there are no perturbative
open string excitations around the stable classical vacuum $\Psi_0$.
This analysis was only carried out for even level scalar fields; it
would be nice to check that a similar result holds for odd-level
fields and for tensor fields of arbitrary rank.

Another more abstract argument that there are no open string states in
the stable vacuum was given by Ellwood, Feng, He and Moeller
 \cite{efhm}.  These authors argued that in the stable vacuum, the
identity state $| I \rangle$ in the SFT star algebra, which satisfies
$ I \star A = A$ for a very general class of string fields
$A$, seems to be an exact state, 
\begin{eqnarray}
| I \rangle
 & = &  \tilde{Q} | \Lambda
\rangle\,. \label{eq:i-exact}
\end{eqnarray} 
If indeed the identity is exact, then it follows
immediately that the cohomology of $\tilde{Q}$ is empty, since
$\tilde{Q}A = 0$ then implies that
\begin{eqnarray}
A & =  & (\tilde{Q} \Lambda)  \star A \nonumber\\
 & = & \tilde{Q} (\Lambda \star A) -\Lambda \star \tilde{Q} A\\
& = & \tilde{Q} (\Lambda \star A) \,. \nonumber
\end{eqnarray}
So to prove that the cohomology of $\tilde{Q}$ is trivial, it suffices
to show that $\tilde{Q} | \Lambda \rangle = | I \rangle$.
While there are some subtleties  involved with the identity
string field, Ellwood {\it et al.} found a very elegant expression for
this field,
\begin{eqnarray}
| I \rangle & = &  
\left( \cdots e^{\frac{1}{8} L_{-16}} e^{\frac{1}{4} L_{-8}}
e^{\frac{1}{2} L_{-4}}\right)
e^{L_{-2}} | 0 \rangle\,.
\end{eqnarray}
(Recall that $| 0 \rangle= b_{-1}| 0_1 \rangle$.)
They then looked numerically for a state $| \Lambda \rangle$
satisfying (\ref{eq:i-exact}).  For example, truncating at level $L = 3$,
\begin{eqnarray}
| I \rangle & = & | 0 \rangle + L_{-2}| 0 \rangle 
+ \cdots\label{eq:identity-2}\\
 & = & | 0 \rangle-b_{-3} c_{1}| 0 \rangle -2b_{-2}c_0| 0 \rangle
 +\frac{1}{2} (\alpha_{-1} \cdot \alpha_{-1})| 0 \rangle 
+ \cdots\nonumber
\end{eqnarray}
while the only candidate for $|\Lambda\rangle$ is
\begin{equation}
| \Lambda \rangle = \alpha \; b_{-2}| 0 \rangle,
\end{equation}
for some constant $\alpha$.  The authors of \cite{efhm} showed that
the state (\ref{eq:identity-2}) is best approximated as exact when
$\alpha \sim 1.12$; for this value, their measure of exactness becomes
\begin{equation}
\frac{\left| \tilde{ Q} |\Lambda \rangle -| I \rangle \right
  |}{| I |}  \rightarrow 0.17,
\end{equation}
which the authors interpreted as a $17\%$ deviation from exactness.
Generalizing this analysis to higher levels, they found at levels 5,
7, and 9, a deviation from exactness of $11\%, 4.5\%$ and $3.5\%$
respectively.  At level 9, for example, the identity field has 118
components, and there are only 43 gauge parameters, so this is a
highly nontrivial check on the exactness of the identity.  Like the
results of \cite{Ellwood-Taylor-spectrum}, these results strongly
support the conclusion that the cohomology of the theory is trivial in
the stable vacuum.  In this case, the result applies to fields  of all
spins and all ghost numbers.

Given that the Witten string field theory seems to have a classical
solution with no perturbative open string excitations, in accordance
with Sen's conjectures, it is quite interesting to ask what the
physics of the vacuum string field theory (\ref{eq:VSFT-action})
should describe.  One natural assumption might be that this theory
should include closed string states in its quantum spectrum.
Unfortunately, addressing this question requires performing
calculations in the quantum theory around the stable vacuum.  Such
calculations are quite difficult (although progress in this direction
has been made in the $p$-adic version of the theory \cite{Minahan}).
Even in the perturbative vacuum, it is difficult to systematically
study closed strings in the quantum string from theory.  We discuss
the question again briefly in the final section.


\section{Further developments}
\label{sec:further}

In this section we review briefly some further developments which we
do not have time to explore in great detail in these lectures.  In
Subsection 7.1 we discuss the pure ghost BRST operator Ansatz of RSZ
(Rastelli, Sen, and Zwiebach) for vacuum string field theory.  In
Subsection 7.2 we discuss ``sliver'' states and related states; these
states are projectors in the SFT star algebra, and are closely related
to D-branes in the RSZ VSFT model.  
These topics will be discussed in further detail in \cite{Taylor-Zwiebach}

\subsection{The vacuum string field theory model of RSZ}

In \cite{RSZ}, Rastelli, Sen, and Zwiebach made an intriguing proposal
regarding the form of Witten's string field theory around the stable
tachyon vacuum.  Since the exact form of the BRST operator $\tilde{Q}$
given by (\ref{eq:VSFT-q}) is not known analytically, and is difficult
to work with numerically, these authors suggested that it might be
possible to ``guess'' an appropriate form for this operator (after
suitable field redefinition), using the properties expected of the
BRST operator in any vacuum.  They suggested a simple
class of BRST operators $\hat{Q}$ which satisfy the properties
(a-c) described in Section 4.2 (actually, they impose the slightly
weaker but still sufficient condition $\int (\hat{Q} \Psi \star \Phi +
(-1)^{G_\Psi} \Psi \star \hat{Q} \Phi)$  instead of condition (b)).
In particular, they propose that after a  field redefinition, the BRST
operator of the string field theory in the classically stable vacuum
should be an operator $\hat{Q}$ expressable purely in terms of ghost
operators.  For example, the simplest operator in the class they
suggest is $\hat{Q} = c_0$, which clearly satisfies $\hat{Q}^2$, and
which also satisfies condition (c) and the weaker form of condition
(b) mentioned above.

The RSZ model of vacuum string field theory has a number of attractive
features.   
\begin{itemize}
\item This model  satisfies all the axioms of string field
theory, and has a BRST operator with vanishing cohomology.  
\item
In the RSZ
model, the equation of motion factorizes into the usual equation of
motion
\begin{equation}
\hat{Q} \Psi_{\rm ghost} + g \Psi_{\rm ghost}\star\Psi_{\rm ghost} = 0
\label{eq:EOM-ghosts}
\end{equation}
for the ghost part of the field, and a projection equation
\begin{equation}
\Psi_{\rm matter} =\Psi_{\rm matter} \star\Psi_{\rm matter} 
\end{equation}
for the matter part of the field, where the full string field is given
by
\begin{equation}
\Psi = \Psi_{\rm ghost} \otimes \Psi_{\rm matter} \,.
\end{equation}
Thus, finding a solution of the equation of motion reduces to the
problem of solving the equation of motion in the ghost sector and
identifying projection operators in the string field star algebra.
It was also recently shown \cite{Hata-Kawano,RSZ-closed,Okuyama} that by
taking the BRST operator $\hat{Q}$ to be given by a ghost insertion
localized at the string midpoint, the ghost equation also has
essentially the form of the projection equation.  Thus, this seems to
be a very natural choice for the BRST operator of the RSZ model.
\item
A number of projection operators have been identified in the string
field star algebra.  These projection operators have many of the
properties desired of D-branes.  We will briefly review some aspects of
these projection operators in the next subsection.
\item
Given the projection operators just mentioned, the ratio of 
tensions between D-branes of different dimensionality can be computed
and has the correct value \cite{RSZ-2}\footnote{
This result was known at the time of the lectures.
There was quite a bit
  of recent 
  work on the problem of computing the exact D-brane tension
  \cite{VSFT-tension}.   A very nice recent
  paper by Okawa \cite{Okawa} resolved the question and demonstrated that
  not only the ratio of tensions, but also the tension of an
  individual brane, is correctly reproduced in the RSZ VSFT theory
  when singularities are correctly controlled.}.
\end{itemize}
Despite the successes of the RSZ model, there are some difficult
technical aspects of this picture.  First, it seems very difficult to
actually prove that this model is related to the VSFT around the
stable vacuum in the Witten model, not least because we lack any
analytic control over the Witten theory.  Second, the RSZ model seems
to have a somewhat singular structure in several respects.  Formally,
the action on any well-behaved Fock space state satisfying the
equation of motion will vanish \cite{Gross-Taylor-II}.  Further, the
natural solutions of the projection equation corresponding to the
matter sector of the equation of motion have rather singular
properties \cite{Moore-Taylor}.  Some of these singular properties are
related to the fact that some of the physics in the RSZ model seems to
have been ``pushed'' into the midpoint of the string.  In the Witten
model, the condition that, for example, $Q^2 = 0$ involves a fairly
subtle anomaly cancellation between the matter and ghost sectors at
the midpoint.  In the RSZ model, the matter and ghost sectors are
essentially decoupled, so that the theory seems to have separate singularities
in each sector, which cancel when the sectors are combined.  These are
all indications of a theory with problematic singularities.
While the Witten theory seems to be free of singularities of
this type, it remains to be seen whether resolving the singularities
of the RSZ model or finding an analytic approach to the Witten theory
will be a more difficult problem to solve.

\subsection{Projection operators in SFT}

From the point of view of the RSZ model of VSFT just discussed,
projection operators in the matter sector of the star algebra play a
crucial role in constructing solutions of the equations of motion.
Such projection operators may also be useful in understanding
solutions in the original Witten theory.  
Quite a bit of work has been
done on constructing and analyzing  projectors in the star
algebra since the RSZ model was originally proposed.  Without
going into the technical details,
we now briefly review some of the important features of matter projectors.

The first matter projector which was explicitly constructed is the
``sliver'' state.  This state was identified in conformal field theory
in \cite{Rastelli-Zwiebach}, and then constructed explicitly using
matter oscillators in \cite{Kostelecky-Potting}.  The sliver state
takes the form of a squeezed state
\begin{equation}
\exp \left[\frac{1}{2}\,
 a^{\dagger} \cdot S \cdot a^{\dagger} \right]| 0 \rangle \,.
\label{eq:sliver-squeezed}
\end{equation}
By requiring that such a state satisfy the projection equation $\Psi
\star \Psi = \Psi$, and by making some further assumptions about the
nature of the state, an explicit formula for the matrix $S$ was found
in \cite{Kostelecky-Potting}.

Projectors like the sliver have many properties which are reminiscent
of D-branes.  This relationship between projection operators and
D-branes is familiar from noncommutative field theory, where
projectors also play the role of D-brane solitons \cite{gms} (for a
review of noncommutative field theory, see \cite{Douglas-Nekrasov}).
In the RSZ model, by tensoring an arbitrary matter projector with a
fixed ghost state satisfying the ghost equation of motion
(\ref{eq:EOM-ghosts}), states corresponding to an arbitrary
configuration of D-branes can be constructed.  Particular projectors
like the sliver can be constructed which are localized in any number
of space-time dimensions, corresponding to the codimension of a
D-brane.  Under gauge transformations, a rank 1 projector can be
rotated into an orthogonal rank 1 projector, so that configurations
containing multiple branes can be constructed as higher rank
projectors formed from the sum of orthogonal rank one projectors
\cite{RSZ-3,Gross-Taylor-I}.  This gives a very suggestive picture of
how arbitrary D-brane configurations can be constructed in string
field theory.  While this picture is quite compelling, however, there
are a number of technical obstacles which make this still a somewhat
incomplete story.  As mentioned above, in the RSZ model, many
singularities appear due to the separation of the matter and ghost
sectors.  In the context of the matter projectors, these singularities
manifest as singular properties of the projectors.  For example, the
sliver state described above has a matrix $S$ which has eigenvalues of
$\pm 1$ for any dimension of D-brane
\cite{Moore-Taylor,RSZ-projectors}.  Such eigenvalues cause the state
to be nonnormalizable elements of the matter Fock space.  In the
Dirichlet directions, this lack of normalizability occurs because the
state is essentially localized to a point and is analogous to a delta
function.  In the Neumann directions, the singularity manifests as a
``breaking'' of the strings composing the D-brane, so that the
functional describing the projector state is a product of a function
of the string configurations on the left and right halves of the
string, with no connection mediated through the midpoint.  These
geometric singularities seem to be generic features of matter
projectors, not just of the sliver state \cite{Schnabl,RSZ-projectors}.  These
singular geometric features are one of the things which makes direct
calculation in the RSZ model somewhat complicated, as all these
singularities must be sensibly regulated.  These singularities do not
seem to appear in the Witten theory, where the BRST operator and
numerically calculated solutions seem to behave smoothly at the string
midpoint.  On the other hand, it may be that further study of the
matter projection operators and their cousins in the ghost sector
which satisfy (\ref{eq:EOM-ghosts}) will lead to analytic progress on
the Witten theory.


\section{Conclusions and open problems} 
\label{sec:open}

The work described in these lectures has brought our understanding of
string field theory to a new level.  We now have fairly conclusive
evidence that open string field theory can successfully describe
distinct vacua with very different geometrical properties, which are
not related to one another through a marginal deformation.  The
resulting picture, in which a complicated set of degrees of freedom
defined primarily through an algebraic structure, can produce
different geometrical backgrounds at different solutions of the
equations of motion, represents an important step beyond perturbative
string theory.  Such an approach, where different backgrounds with
different low-energy degrees of freedom arise from a single underlying
formalism, is clearly necessary to discuss questions of a cosmological
nature in string theory.  It is clearly essential, however, to
generalize from the work described here in which the theory describes
distinct {\it open} string backgrounds, to a formalism where different
{\it closed} string backgrounds appear as solutions to an equation of
motion for a single set  of degrees of freedom.

Clearly, it is an important goal to have a formulation of
string/M-theory in which all the currently understood vacua can arise
in terms of a single well-defined set of degrees of freedom.
It is not yet clear, however, how far it is possible go towards this
goal using the current formulations of string field theory.  It may be
that the correct lesson to take from the work described here is simply
that there {\it are} nonperturbative formulations in which distinct
vacua can be brought together as solutions of a single classical
theory, and that one should search for some deeper fundamental
algebraic formulation where geometry, and even the dimension of
space-time emerge from the fundamental degrees of freedom in the same
way that D-brane geometry emerges from the degrees of freedom of
Witten's open string field theory.  A more conservative scenario,
however, might be that we could perhaps use the current framework of
string field theory, or some limited refinement thereof, to achieve
this goal of providing a universal nonperturbative definition of
string theory and M-theory.  Following this latter scenario, we propose
here a series of questions aimed at continuing the recent developments
in open string field theory as far as possible towards this ultimate
goal.   It is not certain that this research program can be carried to
its conclusion, but it will be very interesting to see how far
open string field theory can go in reproducing important
nonperturbative aspects of string theory.

Some open problems:

\begin{enumerate}
\item[1)] The first important unsolved problem in this area is to find
  an analytic description of the tachyonic vacuum.  Despite several
  years of work on this problem, great success with numerical
  approximations, and some insight from the RSZ vacuum string field
  theory model, we still have no good analytic understanding of the
  difference between the D-brane vacuum and the empty vacuum in
  Witten's open cubic string field theory.  It seems almost 
  unbelievable that there is not some elegant analytic solution to
  this problem.  Finding such an analytic
  solution would almost certainly greatly enhance our understanding of
  this theory, and would probably lead to other significant advances.
\item[2)] Another interesting and important unsolved problem is to
  find, either analytically or numerically, a solution of the Witten
  theory describing {\it two} D25-branes.  If open string field theory
  is truly a background-independent theory, at least in the open
  string sense, it should be just as feasible to go from a vacuum with
  one D-brane to a vacuum with two D-branes as it is to go from a
  vacuum with one D-brane to the empty vacuum (or from the vacuum with
  two D-branes to the vacuum with one D-brane, which is essentially
  the same problem as going from one to none).  Despite some work on
  this problem \cite{Ellwood-Taylor-2}, there is as yet no evidence
  that a double D-brane solution exists for the Witten theory on a
  single D-brane.  Several approaches which have been tried (and will
  be described in more detail in
  \cite{Ellwood-Taylor-2}) include: {\it i}) following a positive mass
  field upward, looking for a stable point; this method seems to fail
  because of gauge-fixing problems---the effective potential often develops
  a singularity before reaching the energy $+ T_{25}$, {\it ii})
  following the intuition of the RSZ model and constructing a gauge
  transform of the original D-brane solution which is
  $\star-$orthogonal to the original D-brane vacuum.  It can be shown
  formally that such a state, when added to the original D-brane
  vacuum gives a new solution with the correct energy for a double
  D-brane; unfortunately, however, we have been unable to identify
  such a state numerically in level truncation.

There are several other problems closely related to the double D-brane
  problem.  One related problem is the problem of studying a D0-brane
  lump solution from the tachyon field on a D1-brane wrapped on a
  small circle.  When the circle is sufficiently small, the mass of
  the D0-brane is larger than that of the wrapped D1-brane.  In this
  case, it seems much more difficult to construct the D0-brane lump
  solution than it is when the D0-brane has mass smaller than the
  D1-brane \cite{Moeller-Zwiebach}.  Another possibly related problem
  is the problem of translating a single D-brane of less than maximal
  dimension in a transverse direction.  It was shown by Sen and
  Zwiebach \cite{Sen-Zwiebach-translate} (in a T-dual picture) that
  after moving a D-brane a finite distance of order of the string
  length in a transverse direction, the level-truncated string field
  theory equations develop a singularity.  Thus, in level truncation
  it does not seem possible to move a D-brane a macroscopic distance
  in a transverse direction\footnote{although this can be done
  formally \cite{Kluson}, it is unclear how the formal solution
  relates to an explicit expression in the oscillator language}.  In
  this case, a toy model \cite{Zwiebach-toy} suggests that the problem
  is that the infinitesimal marginal parameter for the brane
  translation ceases to parameterize the marginal trajectory in field
  space after a finite distance, just as the coordinate $x$ ceases to
  parameterize the circle $x^2 + y^2 = 1$ near $x = 1$.  This is
  similar in spirit to the breakdown of Feynman-Siegel gauge along the
  tachyon potential discussed in section 6.1.

  To show that open string field theory is sufficiently general to
  address arbitrary questions involving different vacua, it is clearly
  necessary to show that the formalism is powerful enough to describe
  multiple brane vacua, the D0-brane lump on an arbitrary radius
  circle, and translated brane vacua.  It is currently unclear whether
  the obstacles to finding these vacua are conceptual or technical.
  It may be that the level-truncation approach is not well-suited to
  finding these vacua.  If this is true, however, we may need a
  clearer mathematical formalism for describing the theory.  There is
  currently some ambiguity in the definition of the theory, in terms
  of precisely which states are allowed in the string field.
  Level-truncation in some sense gives a regularization of, and a
  concrete definition to, the theory.  Without level truncation, we
  would need some more definitive mathematical tools for analyzing
  various features of the theory, such as the other vacua mentioned
  here.
\item[3)] Another open question involves the role that closed strings
  play in open string field theory.  As has been known since the
  earliest days of the subject, closed strings appear as poles in
  perturbative open string scattering amplitudes.  This was shown
  explicitly for Witten's SFT in \cite{fgst}, where it was shown that
  closed string poles arise in the one-loop 2-point function.  If
  Witten's theory is well-defined as a quantum theory, it would follow
  from unitarity that the closed string states should also arise in
  some natural sense as asymptotic states of the quantum open string
  field theory.  It is currently rather unclear, however, whether, and
  if so how, this might be realized.  There are subtleties in the
  quantum formulation of the theory which have never completely been
  resolved \cite{Thorn}.  Both older SFT literature
  \cite{Strominger-closed,Shapiro-Thorn} and recent work
  \cite{Gerasimov-Shatashvili,Moore-Taylor,RSZ-closed,Hashimoto-Itzhaki,
  a-Garousi} have
  suggested ways in which closed strings might be incorporated into
  the open string field theory formalism, but a definitive resolution
  of this question is still not available.  If it is indeed possible
  to encode closed string degrees of freedom in some way in the
  quantum open string field theory, it suggests that one could use the
  Witten formalism in principle to not only compute general closed
  string scattering amplitudes, but perhaps even to address questions
  of closed string vacua.  This is clearly an optimistic scenario, but
  one can imagine that the open string theory might really contain all
  of closed string physics as well as open string physics.  This
  scenario is perhaps not so farfetched, as it really represents
  simply a lifting to the level of string field theory of the AdS/CFT
  story, where the massive as well as the massless modes are included.
  Furthermore, the fact that, as discussed in Section 5.5, the open
  string diagrams precisely cover the moduli space of Riemann surfaces
  with an arbitrary number of handles (and at least one boundary),
  suggests that by shrinking the boundaries to closed strings, one
  might neatly describe all perturbative closed string amplitudes in
  the open string language.  On the other hand, it seems quite
  possible that the closed string sector of the theory is encoded in
  a singular fashion (like the encoding of the D-brane in the RSZ
  VSFT model), so that extracting the closed string physics from the open
  string field theory may involve such complicated manipulations that
  one is better off directly working with a closed string formalism.
  It would certainly be nice to have a clearer picture of how far one
  can go in this direction purely from the open string point of view.
\item[4)] Another obvious, but crucial, question is how this whole
  story can be generalized to superstrings.  The naive Witten cubic
  superstring field theory has technical problems arising from contact
  terms between picture-changing operators
  \cite{Wendt,Greensite-Klinkhamer}.  It has been suggested that these
  problems can be resolved directly in the cubic theory
  \cite{Aref'eva}.  Berkovits has also suggested a new non-polynomial
  string field theory framework which seems to deal successfully with
  the contact term problem, at least in the NS-NS sector
  \cite{Berkovits}.  Some preliminary work indicates that numerical
  calculations on the tachyon condensation problem for the open
  superstring can be carried out in the Berkovits model with analogous
  results to those described here for the bosonic open string,
  although the results to date for the superstring are much more
  limited \cite{superstring}.  It would be nice to have a more
  complete picture for the superstring, and some sense of how issues
  like the closed string question would work in the supersymmetric
  framework.
\item[5)] Perhaps the most important lesson we have learned from the
  body of work discussed in these lectures is that open string field
  theory is a consistent framework in which geometrically distinct
  open string backgrounds can arise as classical solutions of a single
  theory.  A fundamental outstanding problem in string theory is to
  find a framework in which different closed string backgrounds arise
  in a similar fashion from some fixed set of degrees of freedom
  within a single well-defined theory.  In principle, we would hope
  that all the different closed string backgrounds would arise as
  solutions of the equations of motion for the fundamental underlying
  degrees of freedom of string field theory, either by incorporating
  closed strings into the open string field theory framework as
  described above, or by working directly in some formulation of
  closed string field theory.  It is quite challenging to imagine a
  single set of degrees of freedom which would encode, in different
  phases, all the possible string backgrounds we are familiar with.  A
  particularly pressing case is that of M-theory.  In principle, a
  nonperturbative background-independent formulation of type II string
  theory should allow one to take the string coupling to infinity in
  such a way that the fundamental degrees of freedom of the theory are
  still actually at some finite point in their configuration space in
  the limit.  This would lead to the vacuum associated with M-theory
  in flat space-time.  It would be quite remarkable if this can be
  achieved in the framework of string field theory.  Given the
  nontrivial relationship between string fields and low-energy
  effective degrees of freedom, however, such a result cannot be ruled
  out.  If this picture could be successfully implemented, it would
  give a very satisfying understanding of how the complicated network
  of dualities of string and M-theory could be represented in terms of
  a single underlying set of degrees of freedom.
\end{enumerate}

\section*{Acknowledgments}
I would like to thank CECS and the organizers of the School on Quantum
Gravity for their support and hospitality, and for an extremely
enjoyable summer school experience.  Thanks to Erasmo Coletti, Ian
Ellwood, David Gross, Nicolas Moeller, Joe Minahan,
Greg Moore, Leonardo Rastelli,
Martin Schnabl, Ashoke Sen, Jessie Shelton, Ilya Sigalov, and Barton
Zwiebach, for many discussions and collaborations which provided the
material for these lectures.  Thanks also to TASI '01, where some of
this material was presented prior to this School.  Particular thanks
to Barton Zwiebach for suggestions and contributions to these
lecture notes, which have substantial overlap with a more extensive
set of lecture notes based on lectures by Zwiebach and myself at TASI
'01, which will appear presently \cite{Taylor-Zwiebach}.  This work
was supported by the DOE through contract \#DE-FC02-94ER40818.

\end{document}